\documentclass[prd,aps,12pt]{revtex4}
\usepackage{graphicx}
\usepackage{latexsym}
\usepackage{amsmath}
\begin{document}

\title{Phase Structure of 
Anisotropic Antiferromagnetic Heisenberg Model
on Layered Triangular Lattice: \\
Spiral State and Deconfined Spin Liquid}

\author{Kazuya Nakane, Takeshi Kamijo\footnote{Present
address: Department of Applied Physics, Osaka University,
Suita, Osaka, 565-0871 Japan} and Ikuo Ichinose} 

\affiliation{Department of Applied Physics, Nagoya Institute of Technology,
Nagoya, 466-8555 Japan}

\begin{abstract}
In the present paper, we study spin-${1 \over 2}$ 
antiferromagnetic (AF) Heisenberg model on layered anisotropic triangular
lattice and obtain its phase structure.
We use the Schwinger bosons for representing spin operators and
also coherent-state path integral for calculating physical quantities.
Finite-temperature properties of the system are investigated by means
of the numerical Monte-Carlo simulations.
Detailed phase diagram of the system is obtained by calculating internal energy,
specific heat, spin correlation functions, etc.
There are AF N\'eel, paramagnetic and spiral states.
Turning on plaquette term (i.e., the Maxwell term on a lattice) of an
emergent U(1) gauge field that flips a pair of parallel spin-singlet bonds, 
we found that there appears a phase that
is regarded as a deconfined spin-liquid state,
though ``transition" to this phase from the paramagnetic phase
is not of second order but a crossover.
In that phase, the emergent gauge boson is a physical gapless
excitation coupled with spinons.
These results support our previous study on AF Heisenberg model 
on a triangular lattice at vanishing temperature.
\end{abstract}
\maketitle

\section{Introduction}

Study of quantum spin models has a long history.
In particular after the discovery of the high-temperature
superconductors, exotic quantum spin states have been intensively explored.
Among them, spin-liquid state with a deconfined spinon has interested
many theoretical and experimental researches\cite{Sachdev1}.
Recently experiments of the anisotropic triangular antiferromagnet (AF magnet)
Cs$_2$CuCl$_4$ revealed the existence of the incommensurate
spiral order at low temperature ($T$) and also spinon-like excitations
at intermediate $T$\cite{CCC1,CCC2}.
One may think that this material can be a candidate for so-called
Z$_2$ spin liquid\cite{Qi}.
Furthermore very recently, evidences for a spin liquid in 
EtMe$_3$Sb[Pd(dmit)$_2$]$_2$ at very low $T$ were reported\cite{Et}.

In the previous paper\cite{Z2}, we studied frustrated AF Heisenberg model on an
anisotropic triangular lattice in two dimensions (2D) at $T=0$.
We used the Schwinger bosons for representing $s={1 \over 2}$
spin operators, and derived an effective model for low-energy region
assuming existence of a short-range spiral order.
Then low-energy excitations are spinons and an emergent gauge field
with local $Z_2$ gauge symmetry. 
We studied the effective gauge model for the quantum AF Heisenberg model
by means of Monte Carlo (MC) simulations, and obtained phase diagram.
There exist the spiral state, paramagnetic (PM) dimer state and 
the spin-liquid state in the phase diagram.
These phases can be also labeled by the gauge dynamics, i.e.,
Higgs, confinement, and Coulomb phases, respectively.

In the present paper, we shall continue the study in the previous work and 
investigate closely related model, i.e.,
AF Heisenberg model on layered anisotropic triangular lattice.
We are interested mostly in finite-$T$ properties of the model.
By using the Schwinger bosons and CP$^1$ path-integral methods, 
direct application of the MC simulations becomes possible {\em without}
assuming any kind of (short-range) orders.
We shall clarify the phase diagram of the model.
Results obtained in this paper support the study in the previous 
paper\cite{Z2}.

This paper is organized as follows.
In Sec.II, we shall introduce models and give a derivation
of the effective gauge-theory model.
By using the Schwinger bosons, local U(1) gauge symmetry
naturally appears.
Then we discuss possible phases in the frustrated AF spin systems.
Section III is devoted to numerical studies.
We investigated phase structure of the model by calculating 
the internal energy, specific heat, spin correlation functions,
etc.
These numerical calculations show that there exist AF, PM, spiral
and spin-liquid phases.
Detailed study on the critical behaviors between these phases is
given.
Section IV is devoted to summary and discussion.

\section{Quantum AF spin model, Schwinger bosons, 
CP$^1$ representation and gauge theory}
\setcounter{equation}{0}

\begin{figure}
\begin{center}
\includegraphics[width=0.45\hsize]{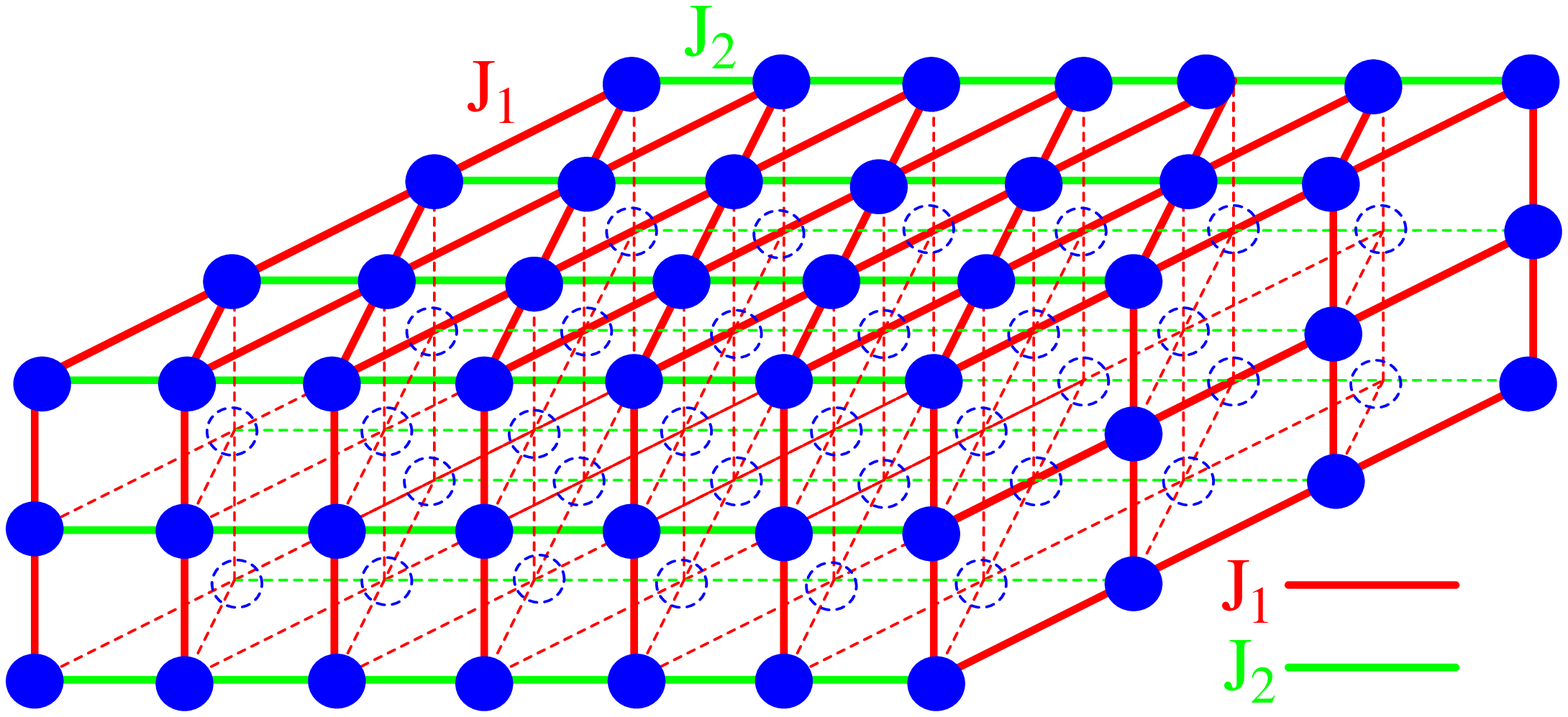}
\includegraphics[width=0.15\hsize]{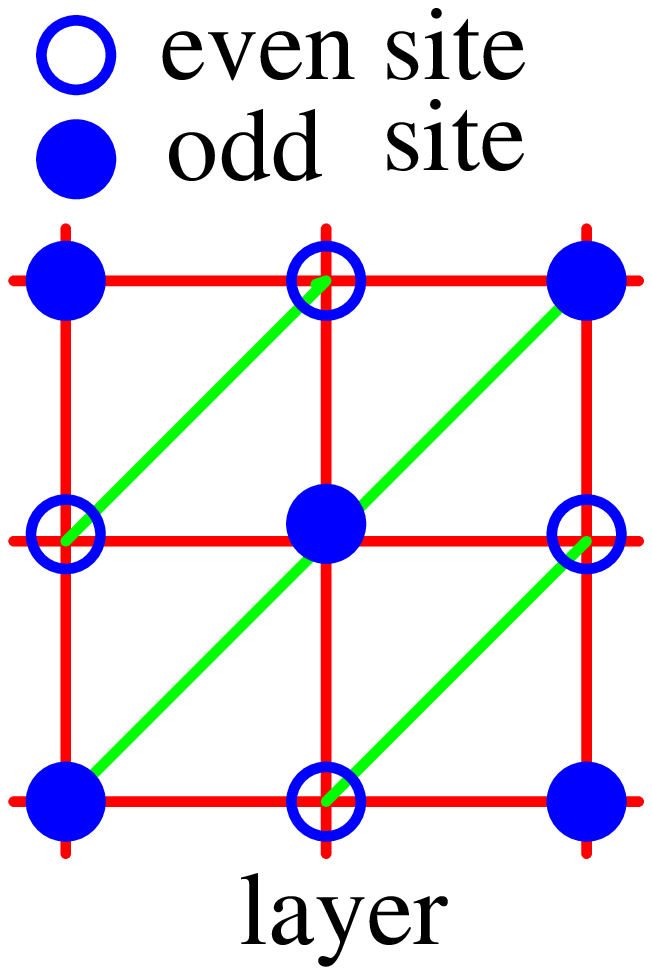}
\end{center}
\caption{Three-dimensional layered lattice on which the spin models 
(\ref{qH}) are defined. 
}
\label{lattice}
\end{figure}

In the present paper we shall study a spin-${1 \over 2}$ anisotropic AF 
Heisenberg model on a layered triangular lattice shown in Fig.\ref{lattice}.
For simplicity, we first consider a 3D cubic lattice and
then add diagonal links in the upper-right direction (1-2 direction) in 2D layers.
Quantum Hamiltonian of the spin system is given as 
\begin{equation}
H=J_1\sum_{x,\mu}\hat{\bf S}_x\cdot\hat{\bf S}_{x+\mu}
+J_2\sum_{x}\hat{\bf S}_x\cdot\hat{\bf S}_{x+1+2},
\label{qH}
\end{equation}
where $\hat{\bf S}_x$ is spin operator at site $x$ and $\mu(=1,2,3)$ is a
direction index and also denotes unit vector in 3D lattice, whereas
$1$ and $2$ are those of the 2D lattice.
Therefore, the $J_1$-term in Eq.(\ref{qH}) is the 3D nearest-neighbor (NN) AF
interaction, whereas the $J_2$-term is the next-nearest-neighbor (NNN) 
AF coupling in 2D layers.
There exists AF N\'eel order for $J_1 \gg J_2$ at low temperature ($T$),
and it is expected that a quantum phase transition takes place 
as $J_2$ is increased.

In this paper, we shall investigate finite-$T$ properties
of the system (\ref{qH}) in detail.
To this end, we employ the Schwinger-boson representation and the 
coherent-path-integral methods\cite{mis}.
By means of these methods, numerical study of the system can be performed
straightforwardly.
In terms of the Schwinger bosons at site $x$, 
$\hat{a}=(\hat{a}_{x\uparrow},\hat{a}_{x\downarrow})^t$
(where ${\cal O}^t$ denotes transpose of the vector/matrix ${\cal O}$),
the spin operator $\hat{\bf S}_x$ is expressed as 
\begin{equation}
\hat{\bf S}_x={1 \over 2}\hat{a}_x^\dagger \vec{\sigma} \hat{a}_x
\label{SB}
\end{equation}
where $\vec{\sigma}$ is the Pauli spin matrices.
As the magnitude of the quantum spin is ${1 \over 2}$,
the physical states of the Schwinger bosons, $|Phys\rangle$,
have to satisfy the 
following constraint at each site $x$,
\begin{equation}
\sum_{\sigma=\uparrow, \downarrow}\hat{a}_{x\sigma}^\dagger\hat{a}_{x\sigma}
|Phys\rangle=|Phys\rangle.
\label{const1}
\end{equation}

We use the coherent-state path integral for the study of the system (\ref{qH})
expressed in terms of the Schwinger bosons.
To this end, we introduce CP$^1$ variables 
$z_x=(z_{x\uparrow},z_{x\downarrow})^t$
corresponding to $\hat{a}_{x\sigma}$,
which satisfy the constraint
\begin{equation}
\sum_{\sigma=\uparrow, \downarrow}\bar{z}_{x\sigma}z_{x\sigma}=1,
\label{CP1}
\end{equation}
as required by Eq.(\ref{const1}).

Then the partition function $Z$ is given by
\begin{eqnarray}
&& Z=\int [Dz]_{\rm CP}\exp \Big[ \int^\beta_0 d\tau A(\tau)\Big],\nonumber \\
&& A(\tau)=-\sum_{x,\sigma}\bar{z}_{x\sigma}
\partial_{\tau}{z}_{x\sigma}
-H(\bar{z},{z}),
\label{Z}
\end{eqnarray}
where $\tau$ is the imaginary time, $\beta=1/(k_BT)$,
$[Dz]$ denotes the path-integral over
CP$^1$ variables and $H(\bar{z},{z})$ is obtained from Eq.(\ref{qH})
by using Eq.(\ref{SB}).
$H(\bar{z},z)$ in Eq.(\ref{Z}) is explicitly given as follows,
\begin{eqnarray}
H(\bar{z},z)&=&{J_1 \over 2}\sum_{x,\mu}|\bar{z}_x{z}_{x+\mu}|^2
+{J_2 \over 2}\sum_{x}|\bar{z}_xz_{x+1+2}|^2 \nonumber  \\
&=&-{J_1 \over 2}\sum_{x,\mu}|\bar{z}_x\tilde{z}_{x+\mu}|^2
-{J_2 \over 2}\sum_{x}|\bar{z}_x\tilde{z}_{x+1+2}|^2
+\mbox{constant},
\label{CPH}
\end{eqnarray}
where $\tilde{z}_x=(\bar{z}_{x\downarrow},-\bar{z}_{x\uparrow})^t$, 
which is nothing but the time-reversal spinor of $z_x$,
and we have used the fact that $z_x$ and $\tilde{z}_x$ are an
orthogonal and complete set of vectors in the CP$^1$ space.
If one tries to numerically study the system (\ref{Z}) by means of
the MC simulations, one immediately encounters difficulties in the
important sampling procedure because the first term in the action 
$A(\tau)$ is pure-imaginary.
For $J_1\gg J_2$,
it is known that by integrating out the CP$^1$ variables $z_x$ at all 
odd (or even) sites of the cubic lattice by assuming a short-range AF order,
resultant action has a quartic form of $z_{x\sigma}$ ($x\in$ even sites) 
and has a lower bound\cite{IM}.
This calculation, however, cannot be applicable for the case 
$J_1 \sim J_2$ that we are interested in.
Therefore we shall take another way to avoid the imaginary term
in $A(\tau)$, i.e., we consider finite-$T$ properties of the
system (\ref{Z}) and ignore the imaginary-time dependence
of variables $z_{x}$.
Study of finite-$T$ properties of the system is not only
interesting itself but also gives an important insight into
low-$T$ properties of the system as it is expected that an ordered
phase at finite $T$ survives at lower $T$'s\cite{aoki}.
In this approximation, the partition function is given as 
\begin{eqnarray}
&& Z=\int [Dz]_{\rm CP}\exp (-S_0),\nonumber \\ 
&& S_0=-\frac{J_1 \beta}{2}\sum _{x,\mu } |\bar{z}_x\tilde{z}_{x+\mu}|^2 
+\frac{J_2 \beta}{2}\sum _{x} |\bar{z}_xz_{x+1+2}|^2 \nonumber \\
&& =-c_1\sum_{x, \mu } |\bar{z}_x\tilde{z}_{x+\mu}|^2 
+d_1\sum_{x} |\bar{z}_xz_{x+1+2}|^2,
\label{Z2}
\end{eqnarray}
where $c_1={J_1\beta \over 2}$ and $d_1={J_2\beta \over 2}$.

For later discussion, it is useful to rewrite the action $S_0$ in 
Eq.(\ref{Z2}) as follows.
We first rename CP$^1$ variables at odd site of the cubic lattice as
\begin{equation}
z_x \rightarrow \tilde{z}_x, \;\; x \in \mbox{odd site},
\label{zodd}
\end{equation}
and introduce a gauge field $U_{x\mu}$ at link $(x,\mu)$ for
the AF spin-pair channel.
Then 
\begin{equation}
S_1=-c'_1\sum_{x, \mu } (\bar{z}_{x+\mu}U_{x\mu}z_x+\mbox{c.c.}) 
+d_1\sum_{x} |\bar{z}_xz_{x+1+2}|^2.
\label{S1}
\end{equation}
One-link integral over the gauge field $U_{x\mu}=e^{i\theta_{x\mu}}$ 
can be performed exactly,
\begin{equation}
\int {d\theta_{x\mu} \over 2\pi}
\exp \Big(c'_1\bar{z}_{x+\mu}U_{x\mu}z_x+\mbox{c.c.}\Big)
=\exp \Big(\log I_0(c'_1|\bar{z}_{x+\mu}z_x|) \Big),
\end{equation}
where $I_0$ is the modified Bessel function.
From the behavior of the modified Bessel function $I_0$, 
relation between the parameters $c_1$ and $c'_1$ is obtained as
follows,
\begin{eqnarray}
c'_1&\sim& 
\left\{
\begin{array}{ll}
c_1 & {\rm for}\ c_1 \gg 1,\\
(c_1)^{1/2}& {\rm for}\ c_1 \ll 1.
\end{array}
\right.
\end{eqnarray}
Action $S_1$ is invariant under the following 
{\em local gauge transformation},
\begin{equation}
z_{x\sigma}\to z_{x\sigma} e^{i\Lambda_x}, \;\;
\bar{z}_{x\sigma}\to \bar{z}_{x\sigma}e^{-i\Lambda_x}, \;\;
U_{x\mu} \to  e^{i\Lambda_{x+\mu}}U_{x\mu}e^{-i\Lambda_x},
\end{equation}
where $\Lambda_x$ is an arbitrary gauge-transformation parameter.
In order to investigate the possibility of the appearance of 
a spin liquid with deconfined spinon excitations, study of the gauge dynamics
and behavior of the gauge field $U_{x\mu}$ is important and necessary.

We also add plaquette term of the gauge field $U_{x\mu}$ to the action
$S_1$ in Eq.(\ref{S1}) as 
\begin{equation}
S_2=-c'_1\sum_{x, \mu } (\bar{z}_{x+\mu}U_{x\mu}z_x+\mbox{c.c.}) 
+d_1\sum_{x} |\bar{z}_xz_{x+1+2}|^2
-c_2\sum_{x,\mu>\nu}U_{x\mu}U_{x+\mu,\nu}\bar{U}_{x+\nu,\mu}
\bar{U}_{x\nu}+\mbox{c.c.},
\label{S2}
\end{equation}
where the last plaquette term is the counterpart
of the Maxwell term of the gauge field $\theta_{x\mu}$ in the continuum, 
and corresponds to the ring-exchange terms of spins like 
$$
(c'_1)^4c_2(\hat{\bf S}_x\cdot\hat{\bf S}_{x+\mu})
(\hat{\bf S}_{x+\nu}\cdot\hat{\bf S}_{x+\mu+\nu})+\cdots,
$$
for the case of small value of $c_2$.
Higher-order terms of $c_2$ correspond to nonlocal interactions
between spins.
It should be noticed that the above plaquette term is defined on
the 3D cubic lattice, and therefore the induced ring-exchange
interaction is three-dimensional.
The gauge field $U_{x\mu}$ is related to the original 
Schwinger-boson operators as 
\begin{equation}
U_{x\mu}\sim 
\left\{
\begin{array}{ll}
\hat{a}_{x+\mu\uparrow}\hat{a}_{x\downarrow}-
\hat{a}_{x+\mu\downarrow}\hat{a}_{x\uparrow}, &
x \in \mbox{odd site} \\
\hat{a}^\dagger_{x+\mu\uparrow}\hat{a}^\dagger_{x\downarrow}
-\hat{a}^\dagger_{x+\mu\downarrow}\hat{a}^\dagger_{x\uparrow}, &
x \in \mbox{even site},
\end{array}
\right.
\label{U}
\end{equation}
i.e., $U_{x\mu}$ corresponds to creation and destruction operators
of spin-singlet bond at sites $x$ and $x+\mu$.
Therefore the $c_2$-terms in the action $S_2$ (\ref{S2})   
flip pairs of parallel nearest-neighbor spin-singlet bonds,
and enhance appearance of the resonating-valence-bond (RVB) liquid\cite{RVB}.

From the previous studies\cite{CPN-2,Z2}, 
phase structure of the quantum spin models
corresponding to $S_2$ in Eq.(\ref{S2}) is expected as follows,
\begin{enumerate}
\item For $d_1=c_2=0$, a phase transition from a paramagnetic state
to the N\'eel state with AF long-range order takes place as $c_1$ 
is increased\cite{AF}.
In a gauge-fixed formalism, the AF N\'eel state corresponds to the state in which
$\langle z_x \rangle \neq 0$ and $\langle U_{x\mu} \rangle \simeq 1$.
\item As the value of $d_1$ is increased in the AF phase, a spiral 
state appears at some critical value of $d_{1c}(c_1)$\cite{spiral}.
In the spiral state, $z_x$ is parameterized as 
$$
z_x={1 \over \sqrt{2}}(e^{i\omega x}v_x+e^{-i\omega x}\tilde{v}_x),
$$
where $\omega$ is a constant,
and the condensation of smoothly varying field $v_x$ takes place,
$\langle v_x \rangle \neq 0$.
\item Furthermore, as the value of $c_2$ is increased, a spin-liquid state
with a deconfined spinon appears.
In the spin-liquid phase, $\langle z_x \rangle=\langle v_x \rangle =0$
and the gauge dynamics of $U_{x\mu}$ is in the Coulomb phase.
Gapless gauge boson appears as a low-energy excitation coupled to spinons.
\end{enumerate}
In the following section, we shall show the results of
study on the phase diagram and physical
properties of the models $S_2$ in Eq.(\ref{S2}) and $S_0$ in Eq.(\ref{Z2}),
which support qualitatively the above expectation.
As mentioned in the introduction, the experiments for frustrated quantum
magnet Cs$_2$CuCl$_4$ observed the spiral state and deconfined 
spin-liquid state\cite{CCC1,CCC2}.
Experimental results suggest a crossover in nature of the excitations
from spin-$1$ spin waves at low energies to deconfined spin-$1/2$ 
spinons at medium to high energies.
Furthermore, EtMe$_3$Sb[Pd(dmit)$_2$]$_2$, which is studied intensively
these days, is closely related to the present model.
Therefore, results in this paper are relevant to these materials.

\section{Numerical studies}
\setcounter{equation}{0}
\subsection{$c_2=0$ case}
In the previous section, we have derived the effective models of 
the CP$^1+$ U(1) gauge variables
from the AF Heisenberg model on layered triangular lattice.
In this section we show results of the numerical study of the models
obtained by means of the MC simulations.
We employed the {\em free boundary condition} in the $1-2$ plane 
as the system may have an incommensurate spiral order with the layered structure.

We first consider the case with $c_2=0$.
We investigated phase structure of the model $S_2$ by calculating
the internal energy $E$ and the specific heat $C$ for various values
of $c'_1$ and $d_1$,
\begin{equation}
E={1 \over L^3}\langle S_2 \rangle, \;\; 
C={1 \over L^3}\langle (S_2-\langle S_2 \rangle )^2\rangle,
\label{EC}
\end{equation}
where $L$ is the system size of the 3D lattice.
In the practical calculation, we employed the local update by 
the standard Metropolis algorithm for the total system with size
$(2+L+2)\times(2+L+2)\times L$ and performed measurement of physical
quantities in the central $L \times L \times L$ subsystem\cite{MC}.

\begin{figure}
\begin{center}
\includegraphics[width=0.5\hsize]{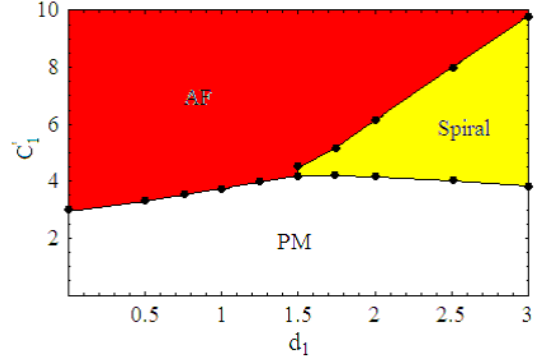}
\end{center}
\caption{Obtained phase diagram for model $S_2$ with $c_2=0$.
There are three phase transition lines, which separate AF, PM and spiral
phases. 
All phase transitions are of second order.
Locations of phase transition lines are determined
by calculations of system size $L=16$.}
\label{F1}
\end{figure}

We have found that $E$ exhibits no anomalous behaviors, whereas
$C$ exhibits singular behaviors 
that indicate existence of second-order phase transitions
as $c'_1$ and $d_1$ are varied.
Observed phase transition lines in the $d_1-c'_1$ plane are
shown in Fig\ref{F1}.
\begin{figure}[htbp]
\begin{center}
\includegraphics[scale=.35]{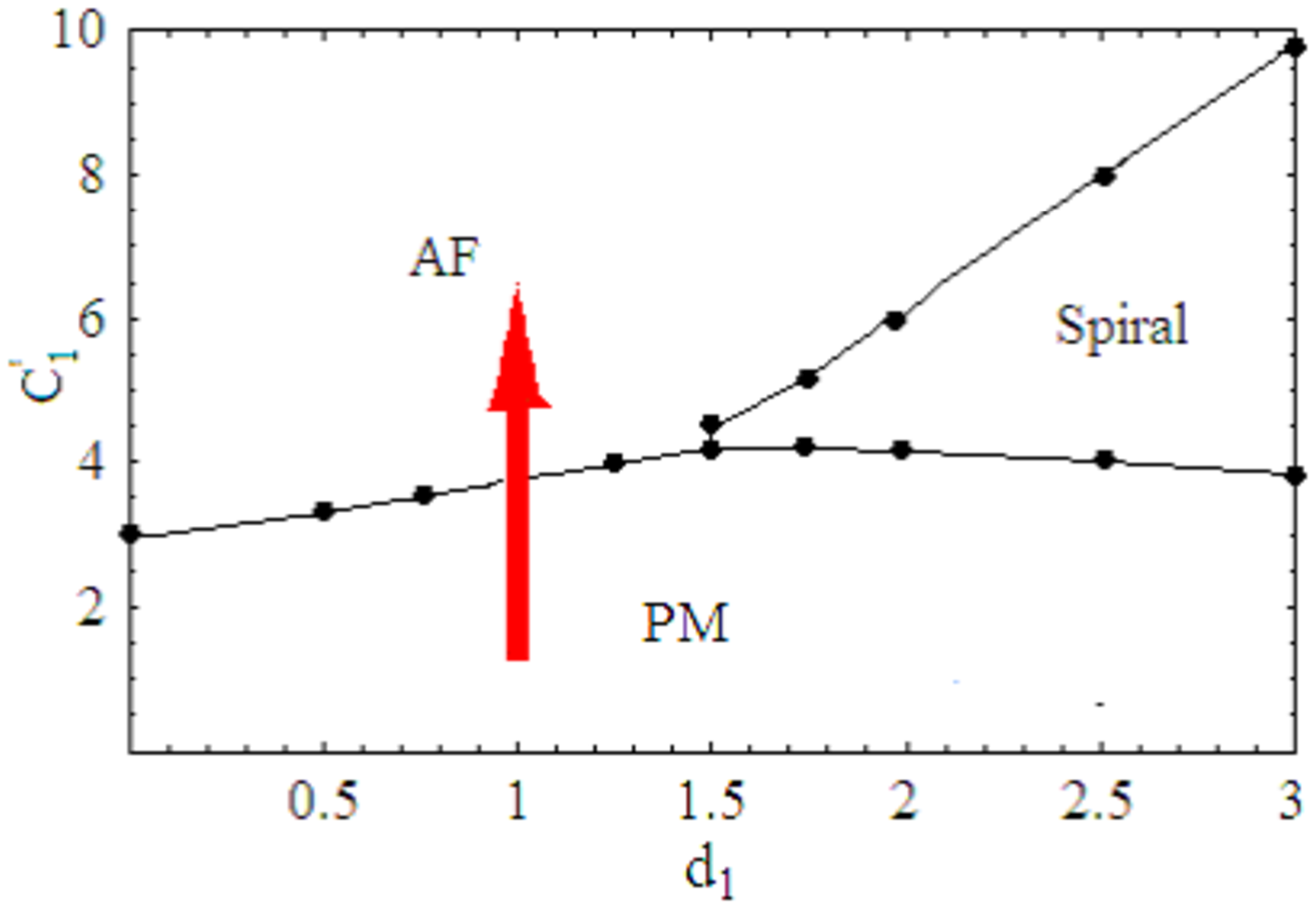} \\
\includegraphics[scale=.35]{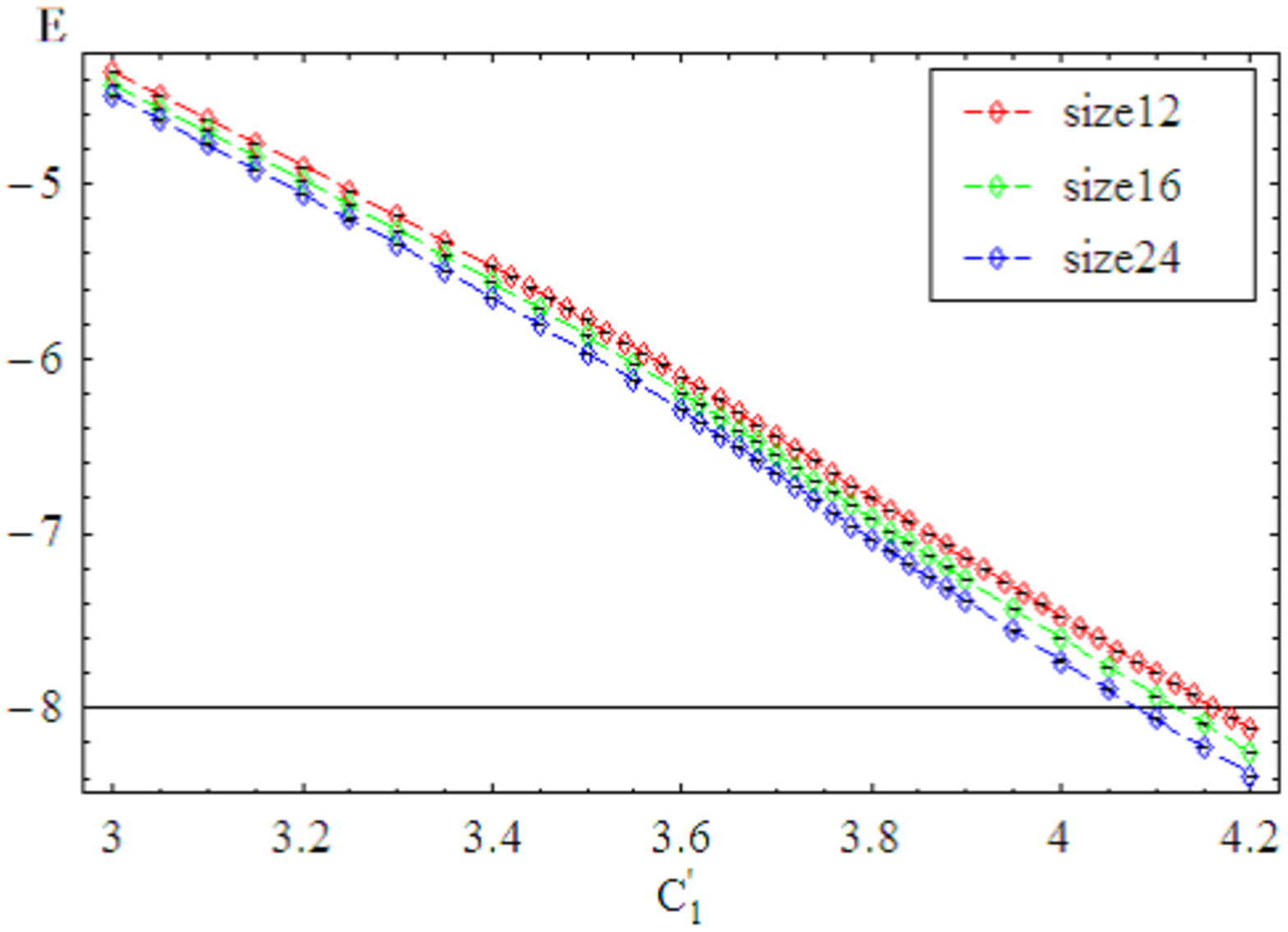}
\includegraphics[scale=.35]{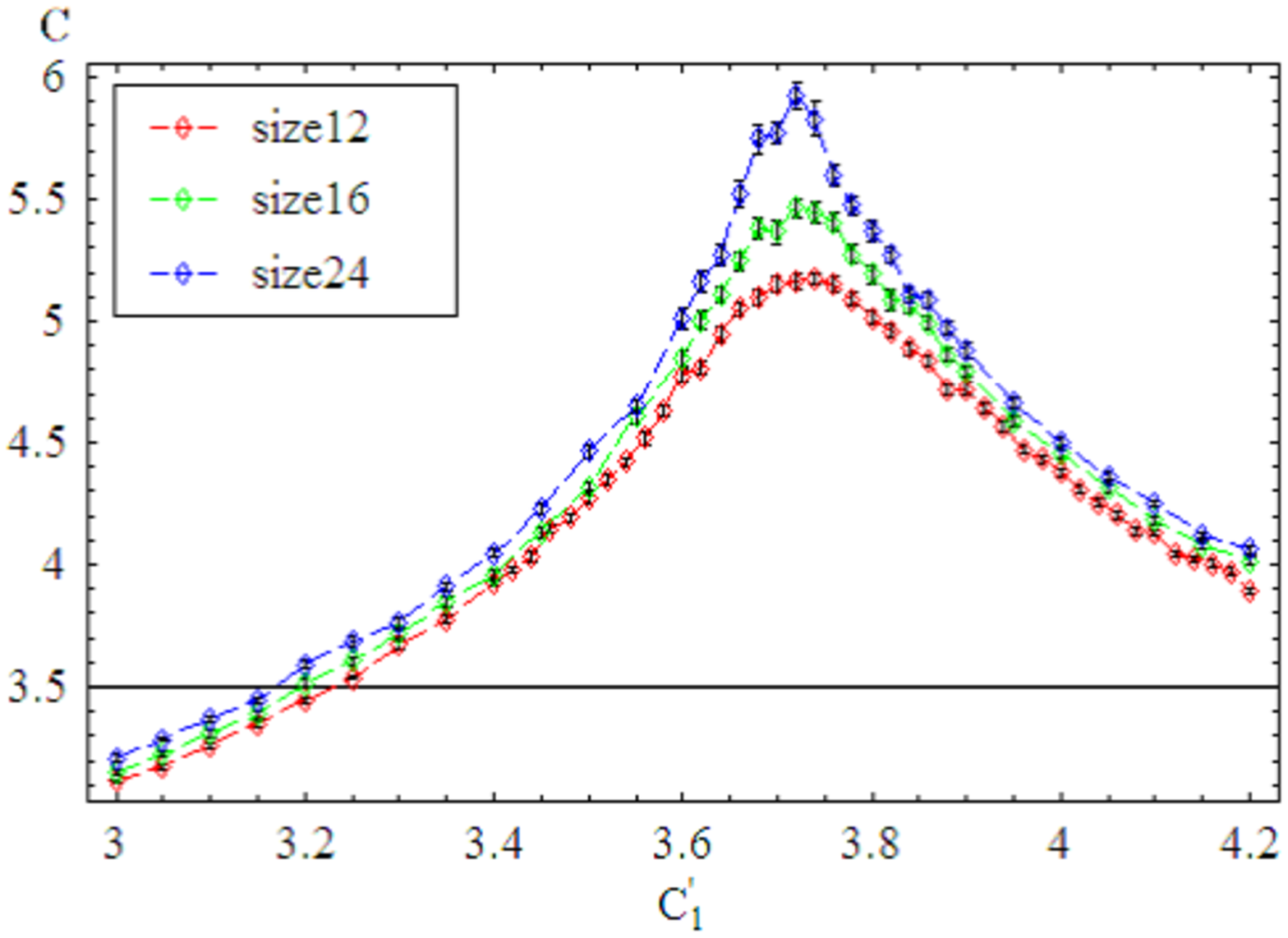}
\end{center}
\caption{Phase transition from PM to AF phases.
$E$ exhibits no anomalous behavior whereas $C$ has a peak indicating
existence of second-order phase transition.}
\label{AFPM}
\end{figure}
\begin{figure}[htbp]
\begin{center}
\includegraphics[width=0.4\hsize]{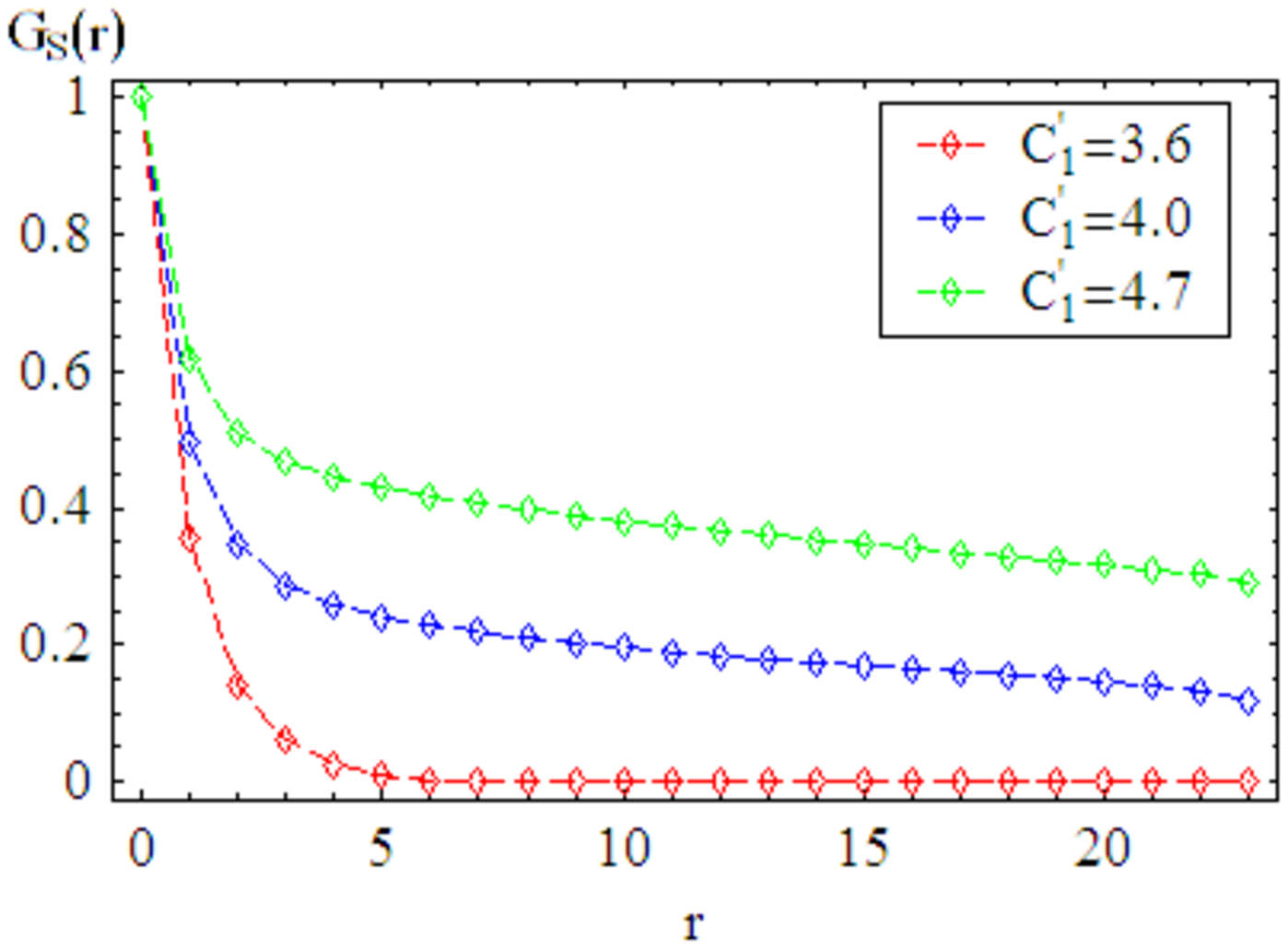}
\includegraphics[width=0.4\hsize]{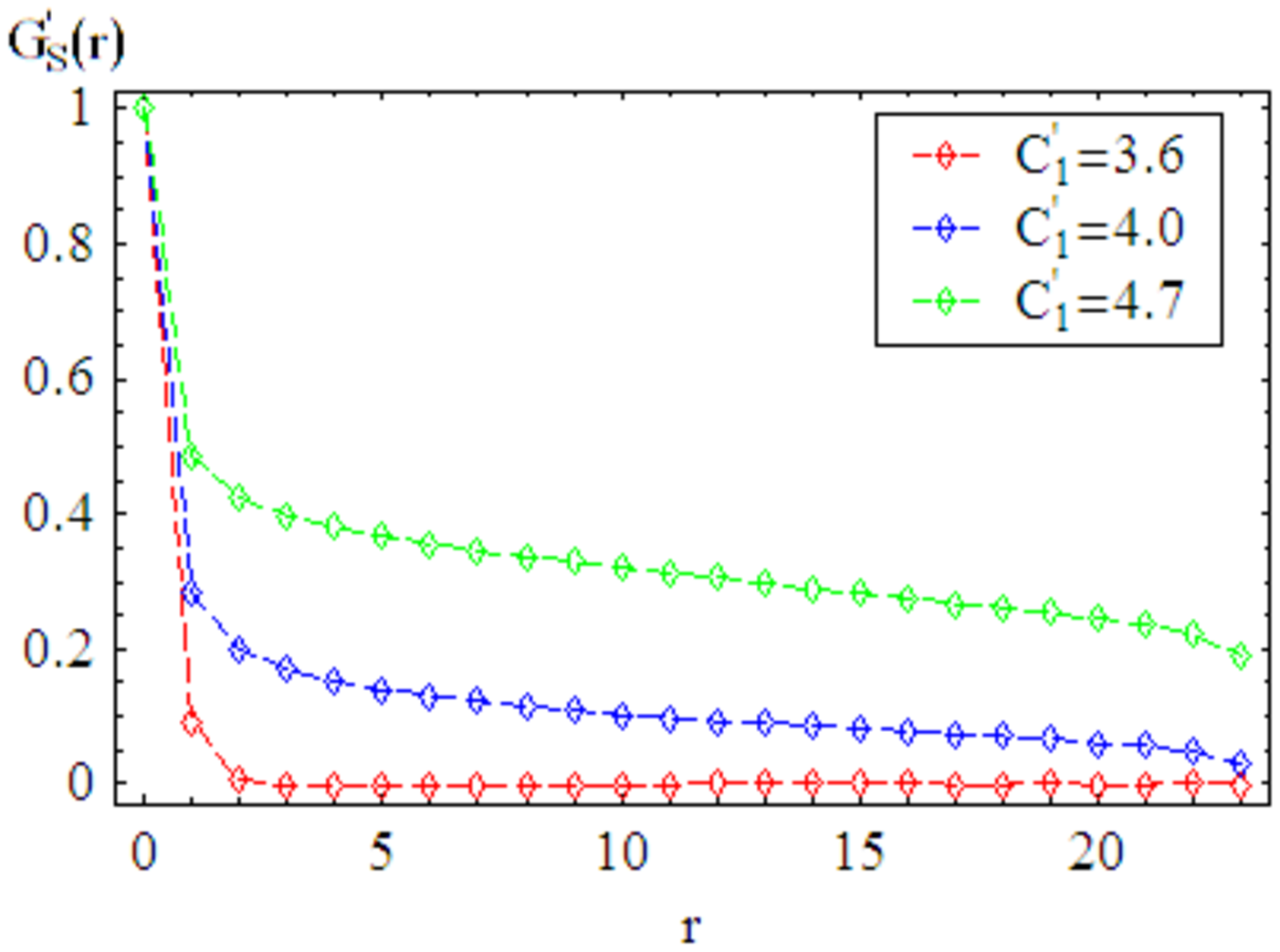}
\end{center}
\caption{Spin correlation functions for $d_1=1.0$.
}
\label{spin1}
\end{figure}

We first consider the PM-AF phase transition.
In Fig.\ref{AFPM}, we show $E$ and $C$ as a function of $c'_1$ for $d_1=1.0$. 
It is obvious that $E$ exhibits no anomalous behavior whereas $C$
has a peak at $c'_1 \simeq 3.75$ and the peak develops as the system size
is increased.
This behavior of $C$ indicates a second-order phase transition at
$c'_{1} \simeq 3.75$.
In order to verify existence of the phase transition and understand physical
meaning of each phase, 
we investigated correlation functions of spins that are given as follows,
\begin{equation}
G_{\rm S}(r)={1 \over 2}\sum_{j=1,2}
\langle {\bf n}_x\cdot{\bf n}_{x+jr}\rangle, \;\;\;
G'_{\rm S}(r)=\langle {\bf n}_x\cdot{\bf n}_{x+(1+2)r}\rangle,
\label{Scor}
\end{equation}
where ${\bf n}_x=(\bar{z}_x\vec{\sigma}z_x)$.
Numerically obtained results are shown in Fig.\ref{spin1}.
At $c'_1=3.6$, the correlation functions have no long-range order (LRO).
On the other hand at $c'_1=4.0, \; 4.7$, they exhibit AF LRO.
(Please recall that we have changed variables 
$z_x \rightarrow \tilde{z}_x,\;\ x \in \mbox{odd site}$.)
From this result, we conclude that transition from the PM to
AF phases takes place at $c'_1 \simeq 3.75$.

\begin{figure}[htbp]
\begin{center}
\includegraphics[scale=.35]{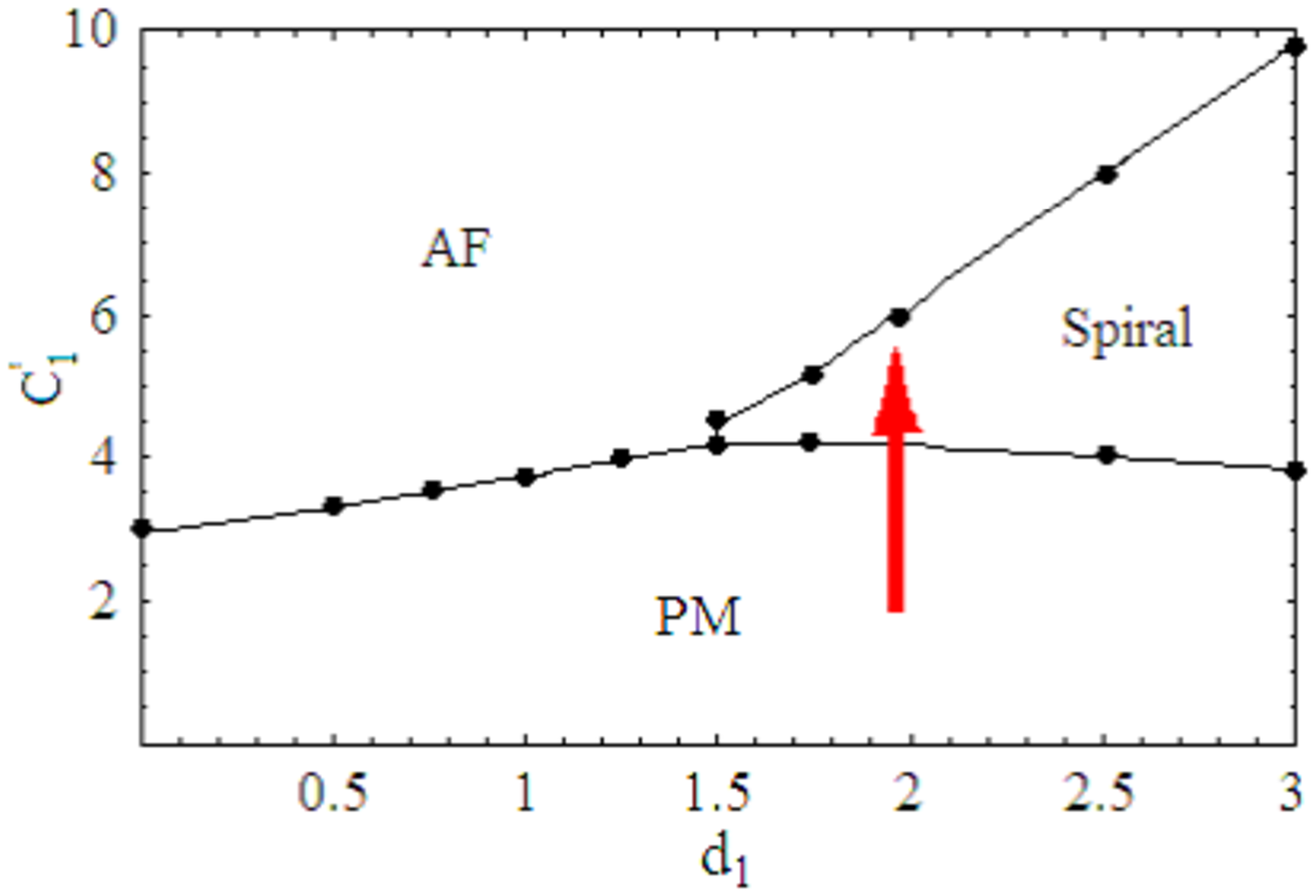}
\includegraphics[scale=.45]{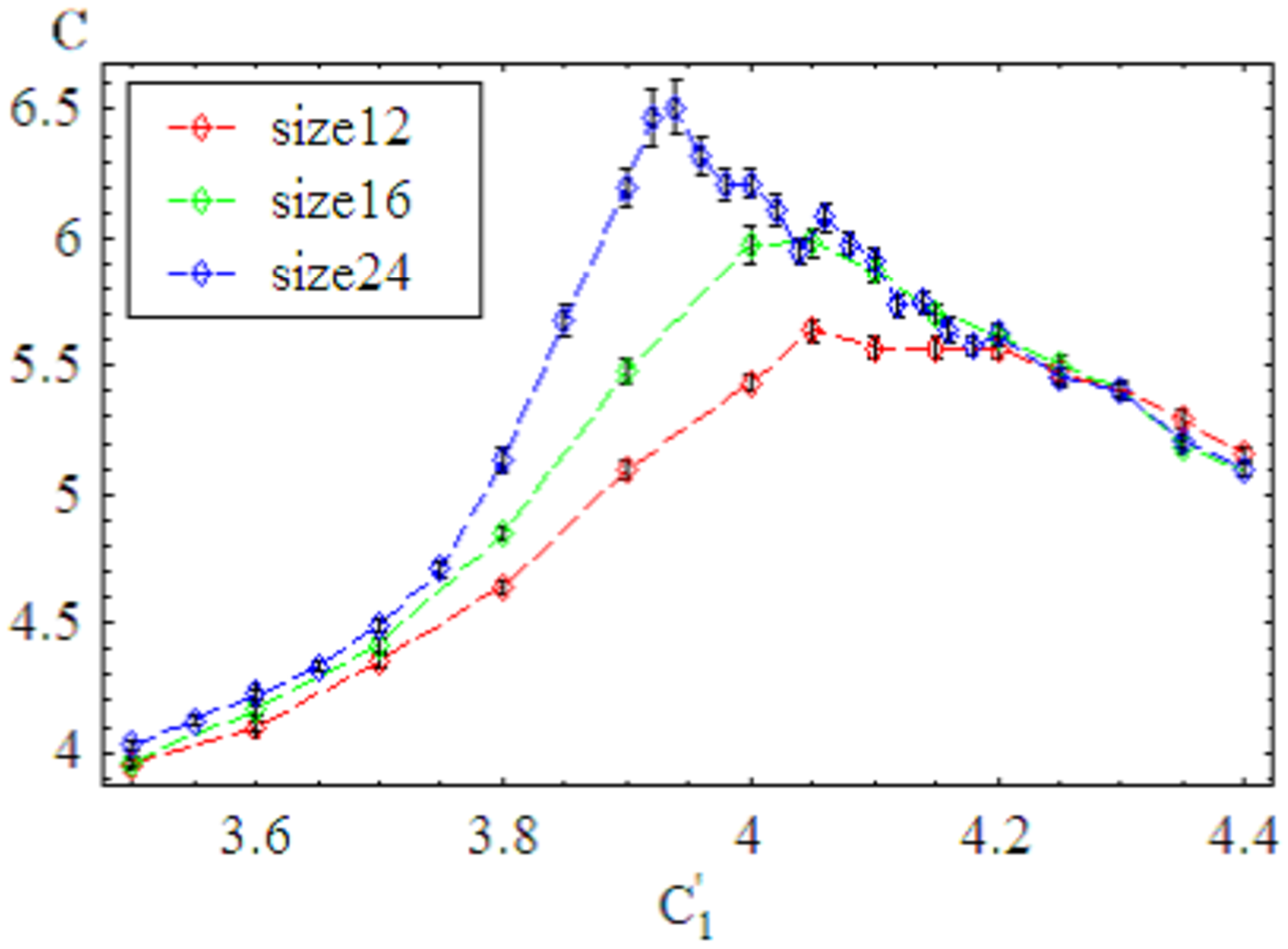}
\end{center}
\caption{Phase transition from PM to spiral states.
Total specific heat $C$ has a peak that develops as $L$ is
increased.}
\label{F2}
\end{figure}

\begin{figure}[htbp]
\begin{center}
\includegraphics[scale=.4]{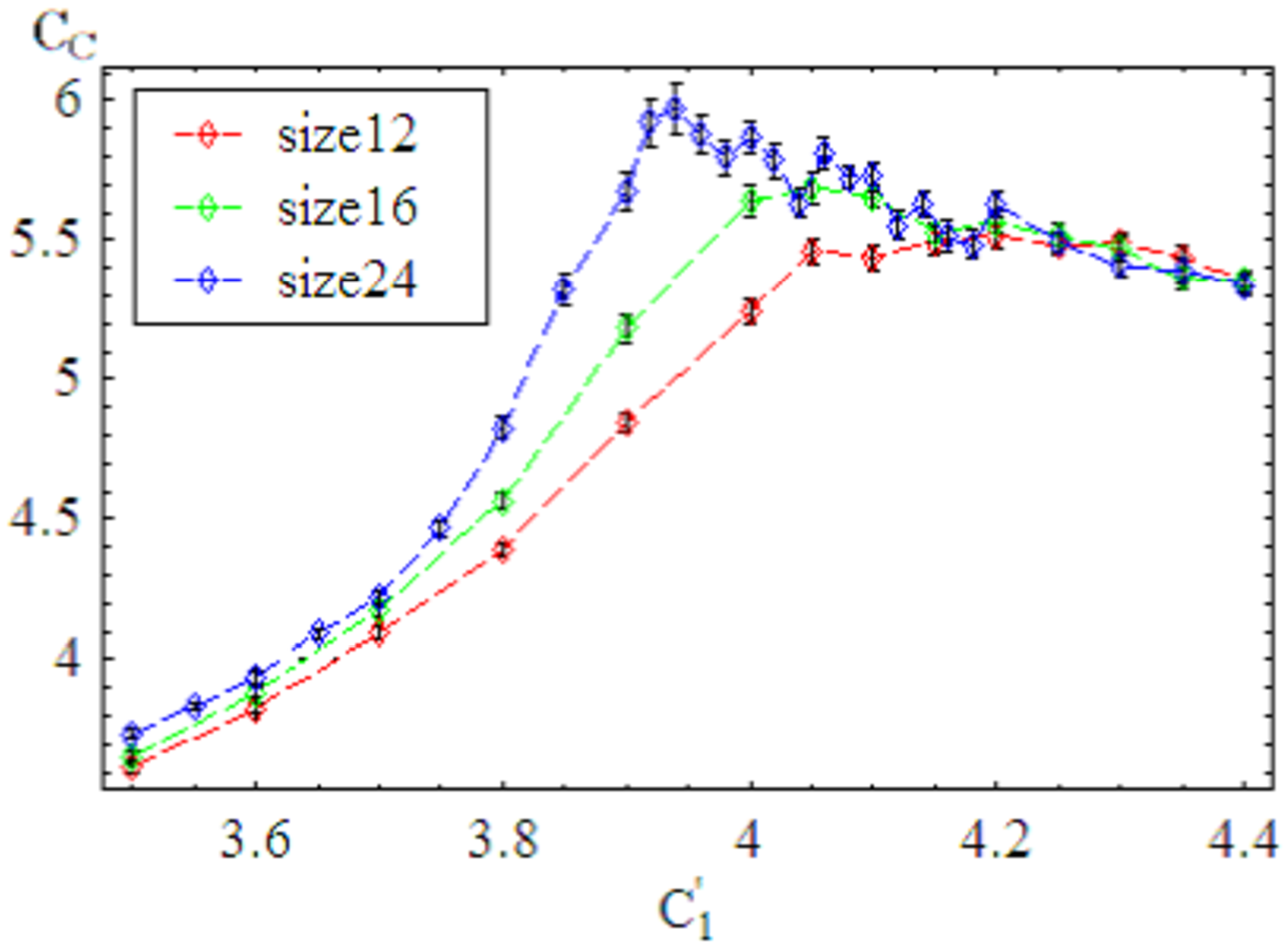}
\includegraphics[scale=.4]{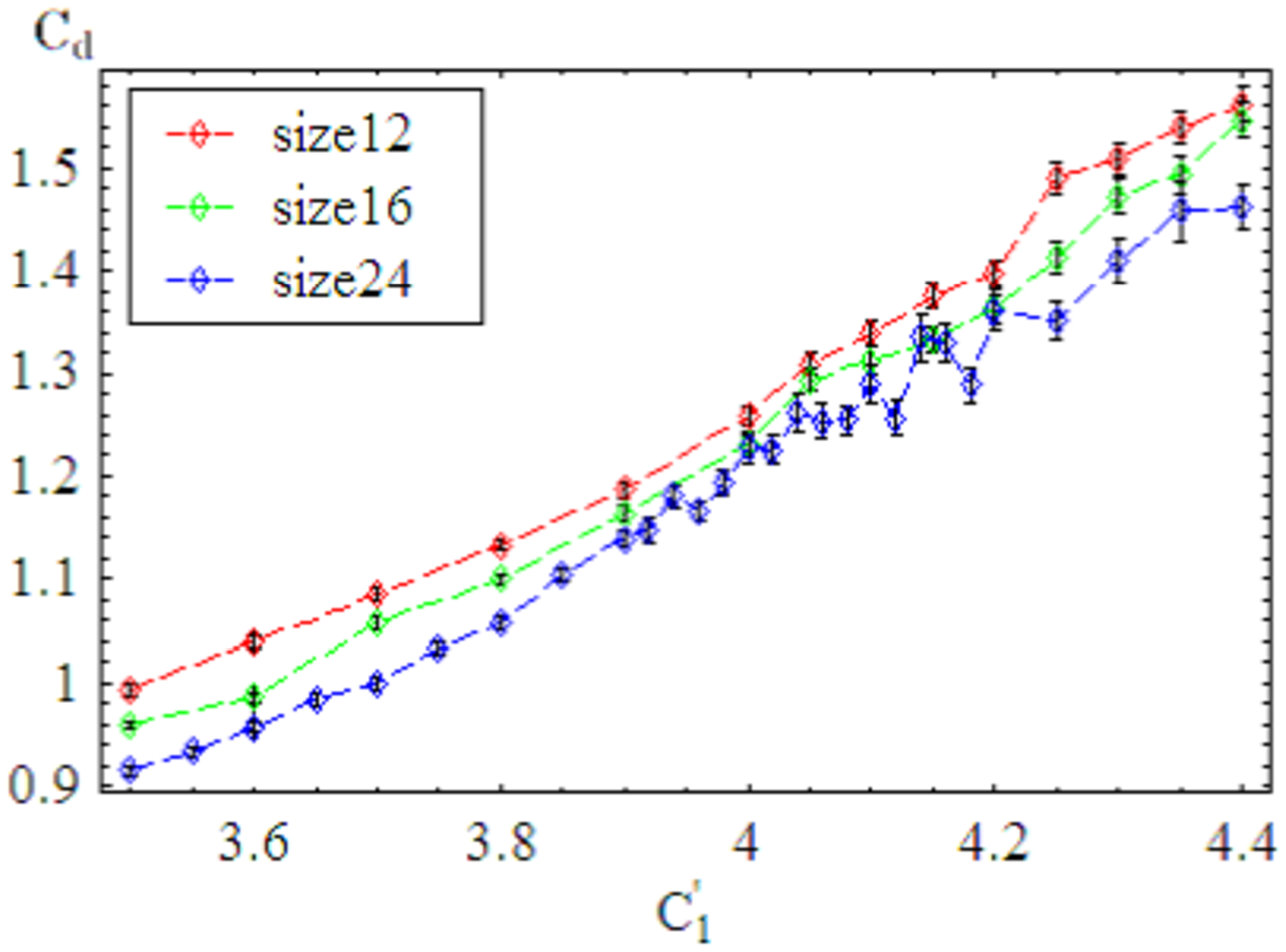}
\end{center}
\caption{Specific heat of each term in action.
Results show that $c_1$-term in action fluctuates strongly
at phase transition but $d_1$-term does not.}
\label{C_each_PM_SP}
\end{figure}

We turn to the phase transition from the PM to spiral states
as shown in Fig.\ref{F2}.
For $d_1=2.0$,
calculation of the total specific heat $C$ as a function of $c'_1$
is shown in Fig.\ref{F2}.
We also measured the specific heat of each term of
the action, which is defined similarly to $C$ in Eq.(\ref{EC}),
in order to see the physical meaning of the phase transition.
\begin{eqnarray}
C_{\rm c}&=& {1 \over L^3}\langle (S_c-\langle S_c\rangle)^2\rangle,
\nonumber  \\
C_{\rm d}&=& {1 \over L^3}\langle (S_d-\langle S_d\rangle)^2\rangle,
\label{CcCd}
\end{eqnarray}
where
\begin{eqnarray}
S_c&=&-c'_1\sum_{x, \mu } (\bar{z}_{x+\mu}U_{x\mu}z_x+\mbox{c.c.}) \nonumber  \\
S_d&=&d_1\sum_{x} |\bar{z}_xz_{x+1+2}|^2.
\end{eqnarray}
See Fig.\ref{C_each_PM_SP}.
From these results, it is obvious that a second-order phase transition 
from the PM to spiral states takes place at $c'_1\simeq 3.9$.
It is interesting to see that the $c'_1$-term of the action tends to
fluctuate strongly at the phase transition point but the $d_1$-term does not.

\begin{figure}[tp]
\begin{center}
\includegraphics[scale=.4]{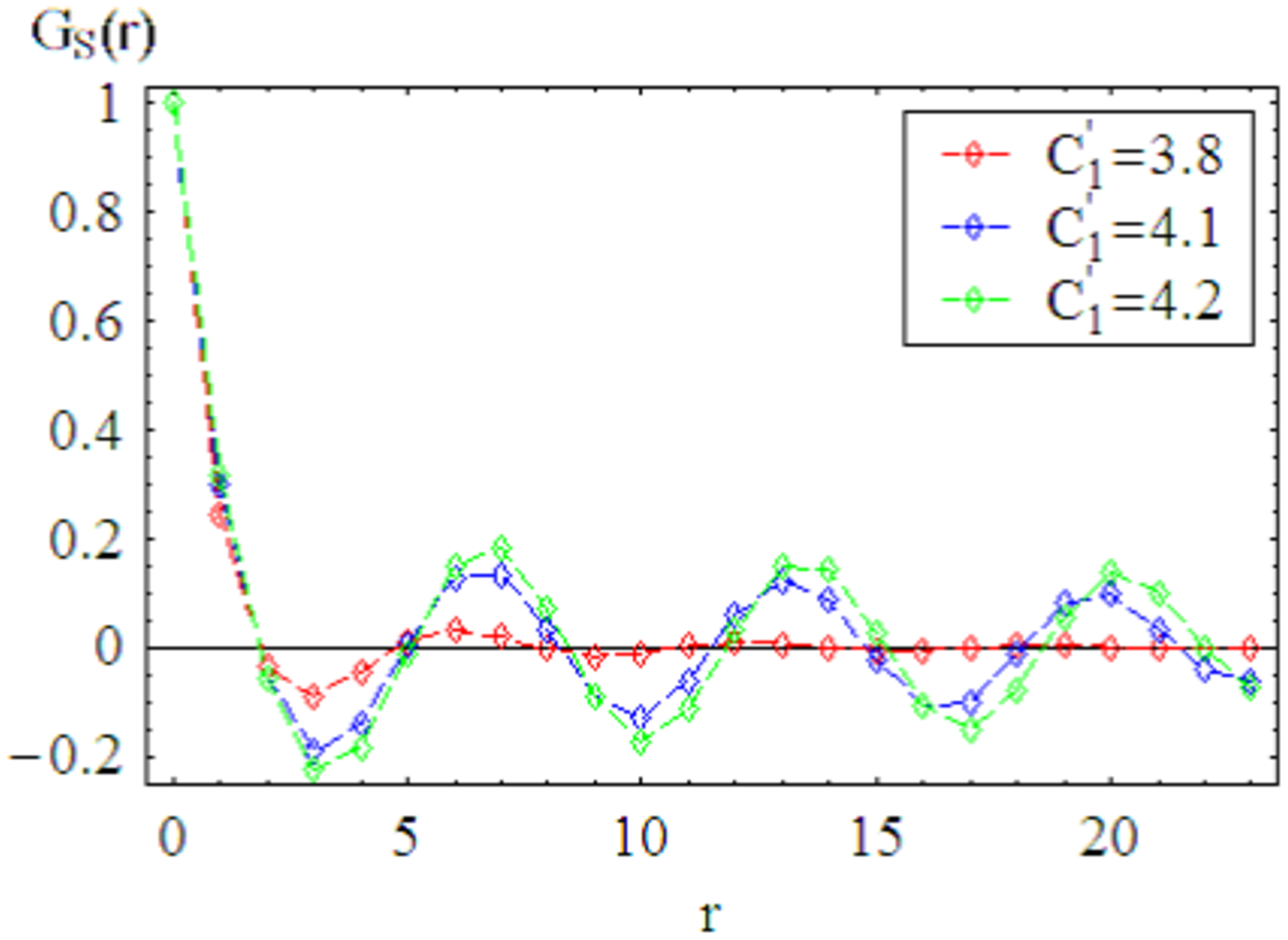}
\includegraphics[scale=.4]{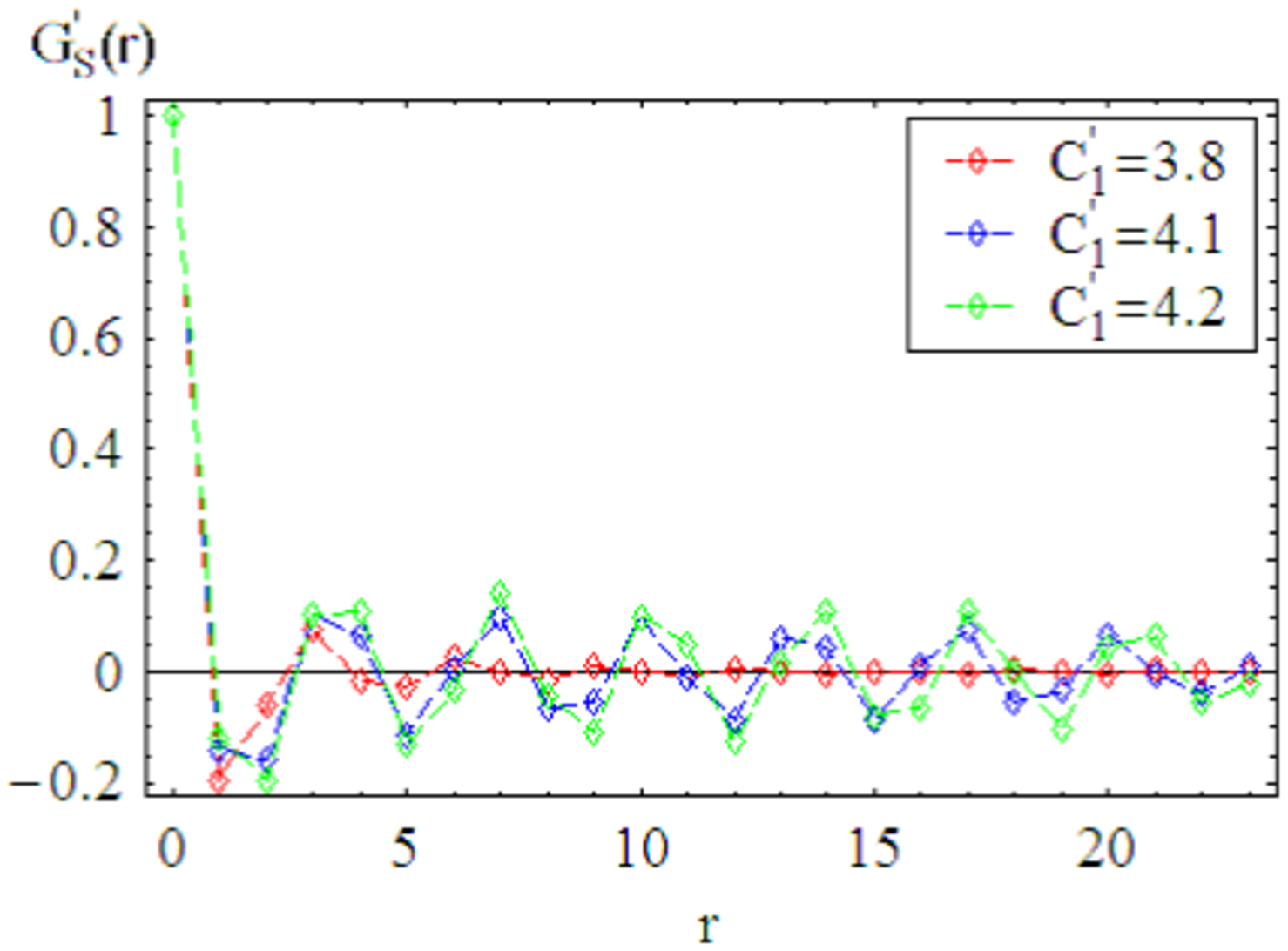}
\end{center}
\caption{Spin correlations in PM (tilted-dimer) and spiral states.
For $c'_1=4.1$ and $4.2$, there is a LRO close to $120^0$-N\'eel
order. Please remember that direction of odd-site spins has been reversed.}
\label{spin_PM_SP}
\end{figure}

\begin{figure}[htbp]
\begin{center}
\includegraphics[scale=.45]{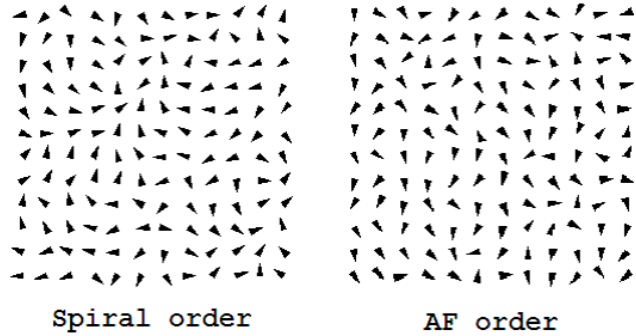}
\end{center}
\vspace{-1cm}
\caption{Snapshots of spin configuration in spiral (left) and 
AF (right) states.
In the snapshots, direction of spins at all odd sites is inverted.
Arrows indicate direction of spins projected into the 1-2 plane. }
\label{snap}
\end{figure}

We measured the spin correlations at $c'_1=3.8, \; 4.1$ and $4.2$.
The results are shown in Fig.\ref{spin_PM_SP}.
It is obvious that at $c'_1=3.8$ the spin does not have a LRO,
whereas at $c'_1=4.1, \; 4.2$ it has a spiral LRO.
One should notice, however, that at $c'_1=3.8$ the spin correlation
has a short-range spiral order and therefore we call this
``phase" a {\em tilted-dimer state}, though there is no sharp
phase boundary between the ordinary PM (at $d_1\ll 1$) and tilted-dimer state.
This observation supports our previous study of the AF magnets
on anisotropic triangular lattice assuming short-range spiral order\cite{Z2}.
We also measured the spin correlation in the inter-layer direction
in the spiral state, and found that it has an ordinary AF correlation as it is
expected.
It is interesting to see a snapshot of spin configurations
in the spiral and AF states.
See Fig.\ref{snap}.

We calculated the specific heat and spin correlations for various 
values of the parameters $c'_1$ and $d_1$ and have obtained 
phase transition line that separates the PM and spiral phases.


\begin{figure}[tp]
\begin{center}
\includegraphics[scale=.4]{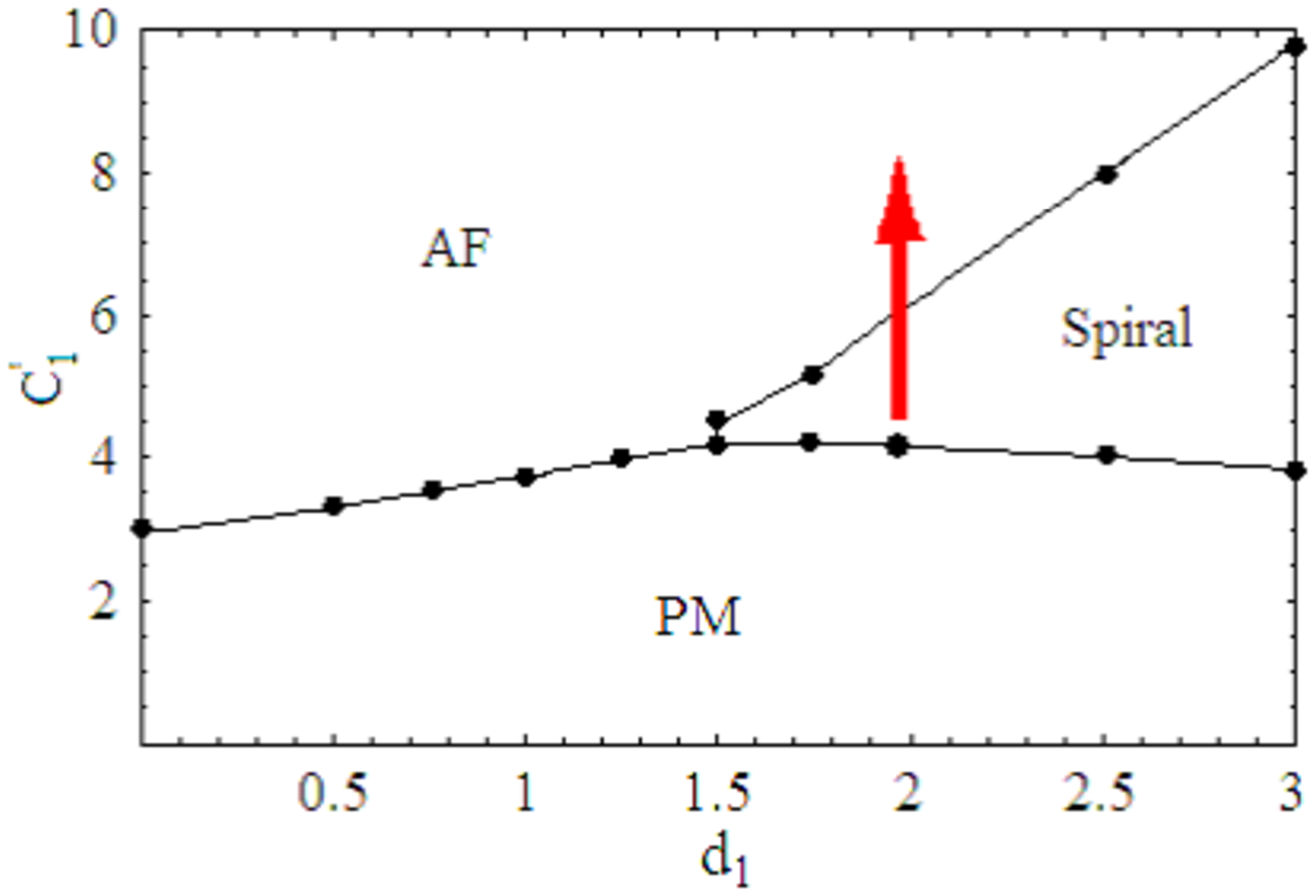}
\includegraphics[scale=.4]{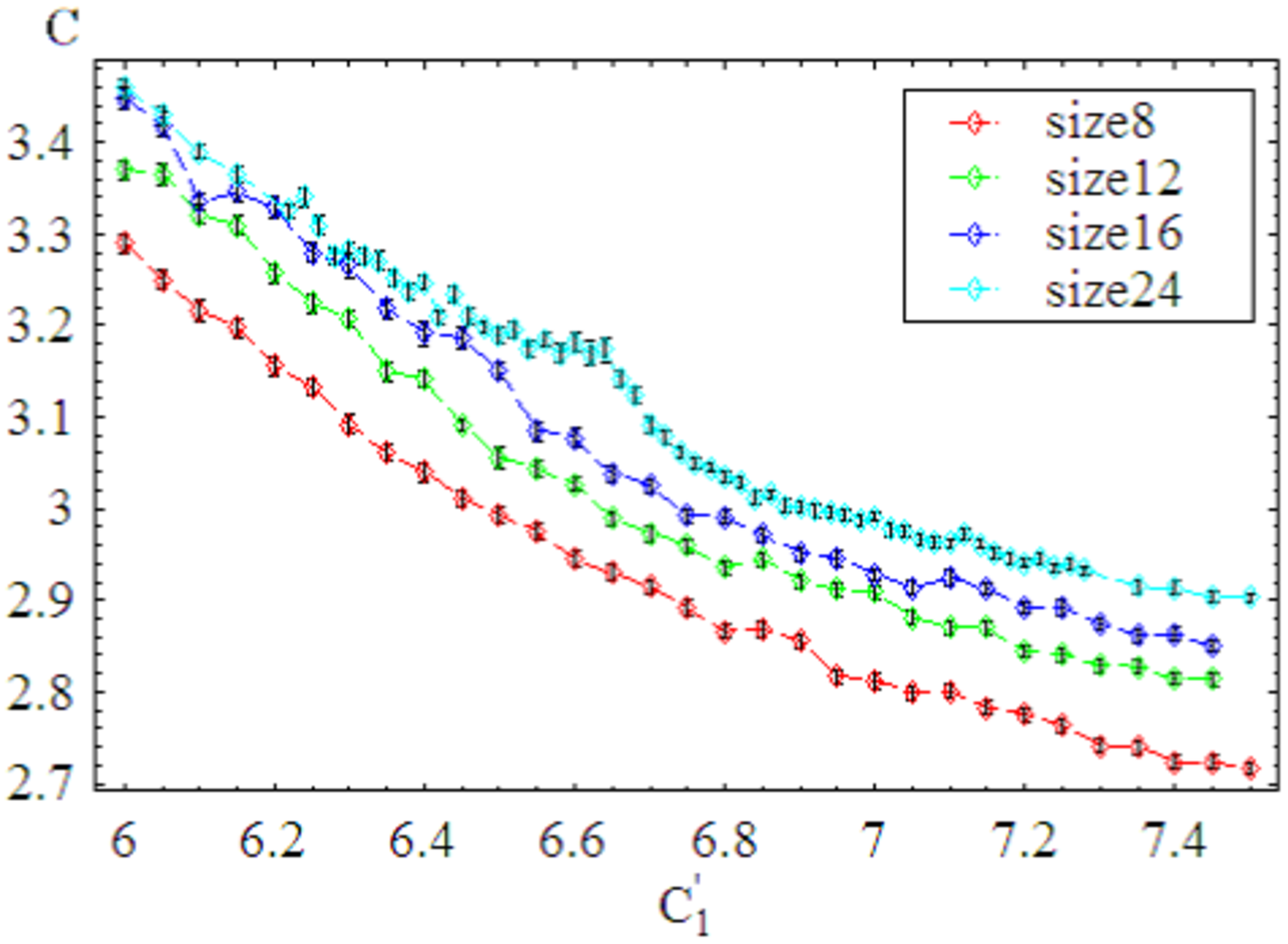}
\end{center}
\caption{Phase transition from the spiral to AF phases.
Total specific heat $C$ for spiral to AF phase transition
exhibits only weak anomalous behavior at phase transition.}
\label{F3}
\end{figure}

\begin{figure}[htbp]
\begin{center}
\includegraphics[scale=.35]{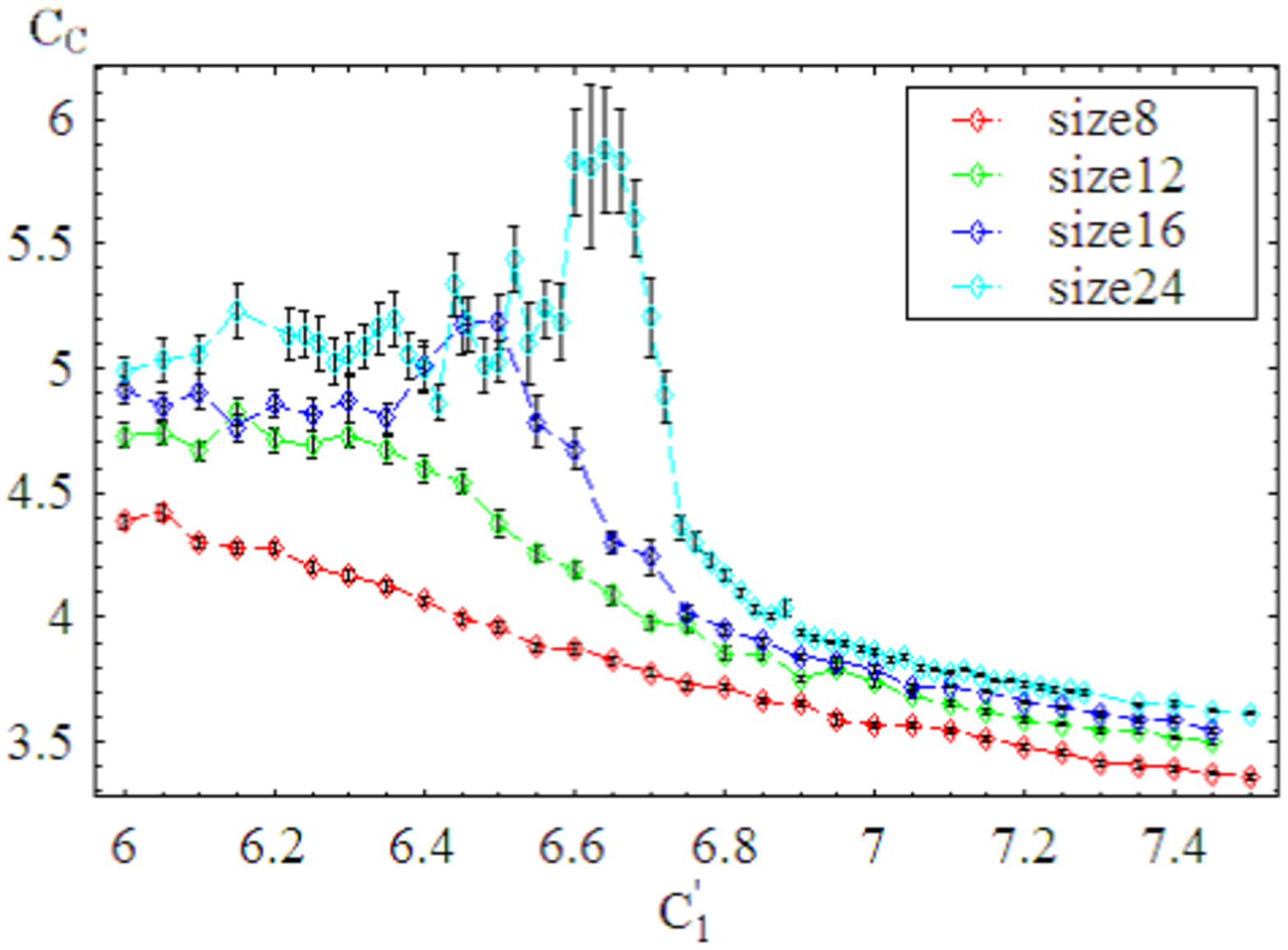}
\includegraphics[scale=.35]{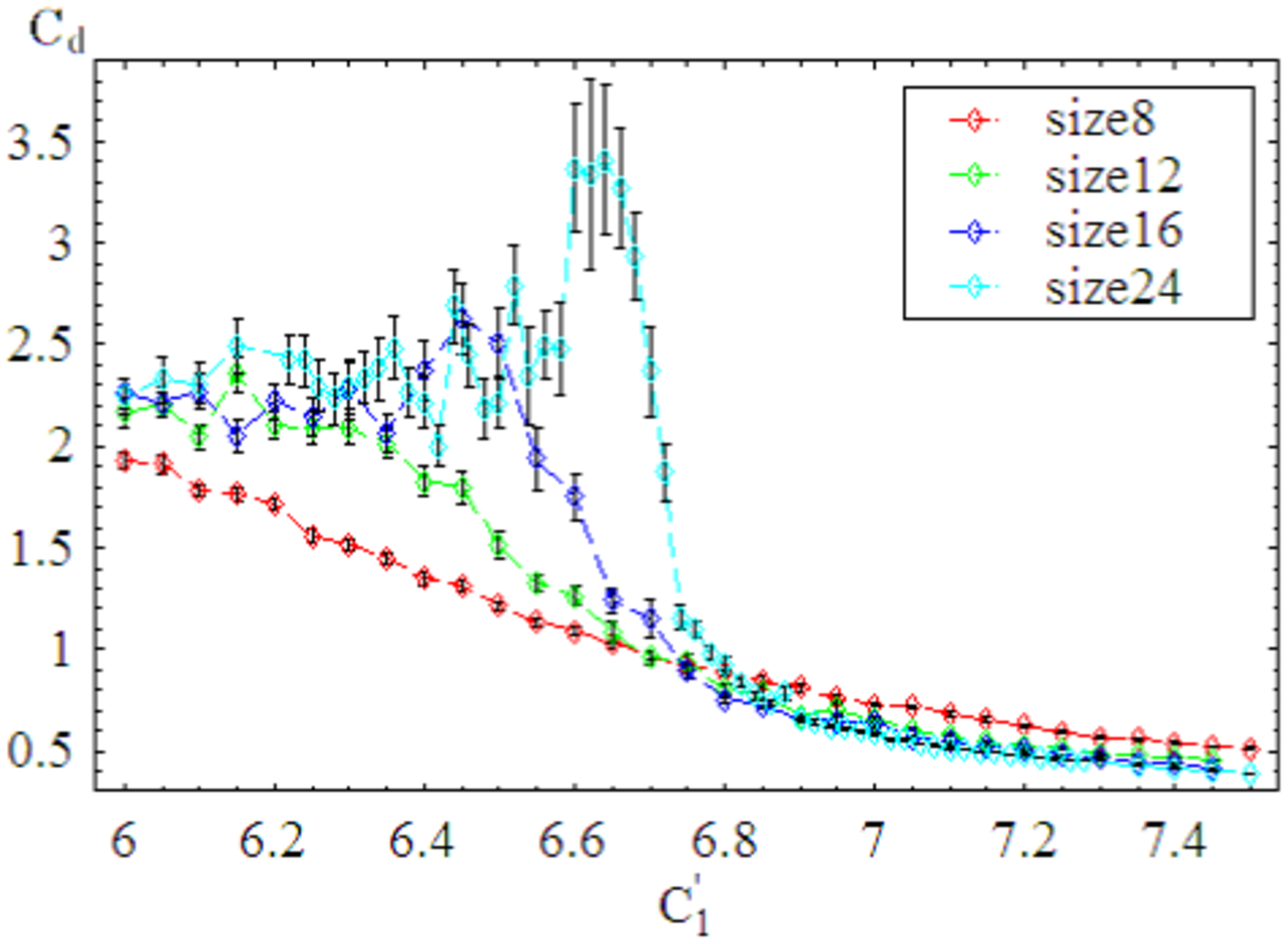}
\end{center}
\caption{Specific heat of each term for spiral to AF phases transition.}
\label{C_each_SP_AF}
\end{figure}

Finally let us turn to the spiral-AF phase transition
(see Fig.\ref{F3}.).
We show calculations of $C$ and the specific heat of $c'_1$ and
$d_1$-terms.
See Figs.\ref{F3} and \ref{C_each_SP_AF}.
It is obvious that the total specific heat $C$ exhibits only very
weak anomalous behavior but $C_{\rm c}$ and $C_{\rm d}$ both show sharp
peak at $c'_1\sim 6.7$ as the system size is increased.
From this result, we conclude that the transition from the spiral to
AF phases is of second order.


\begin{figure}[htbp]
\begin{center}
\includegraphics[scale=.3]{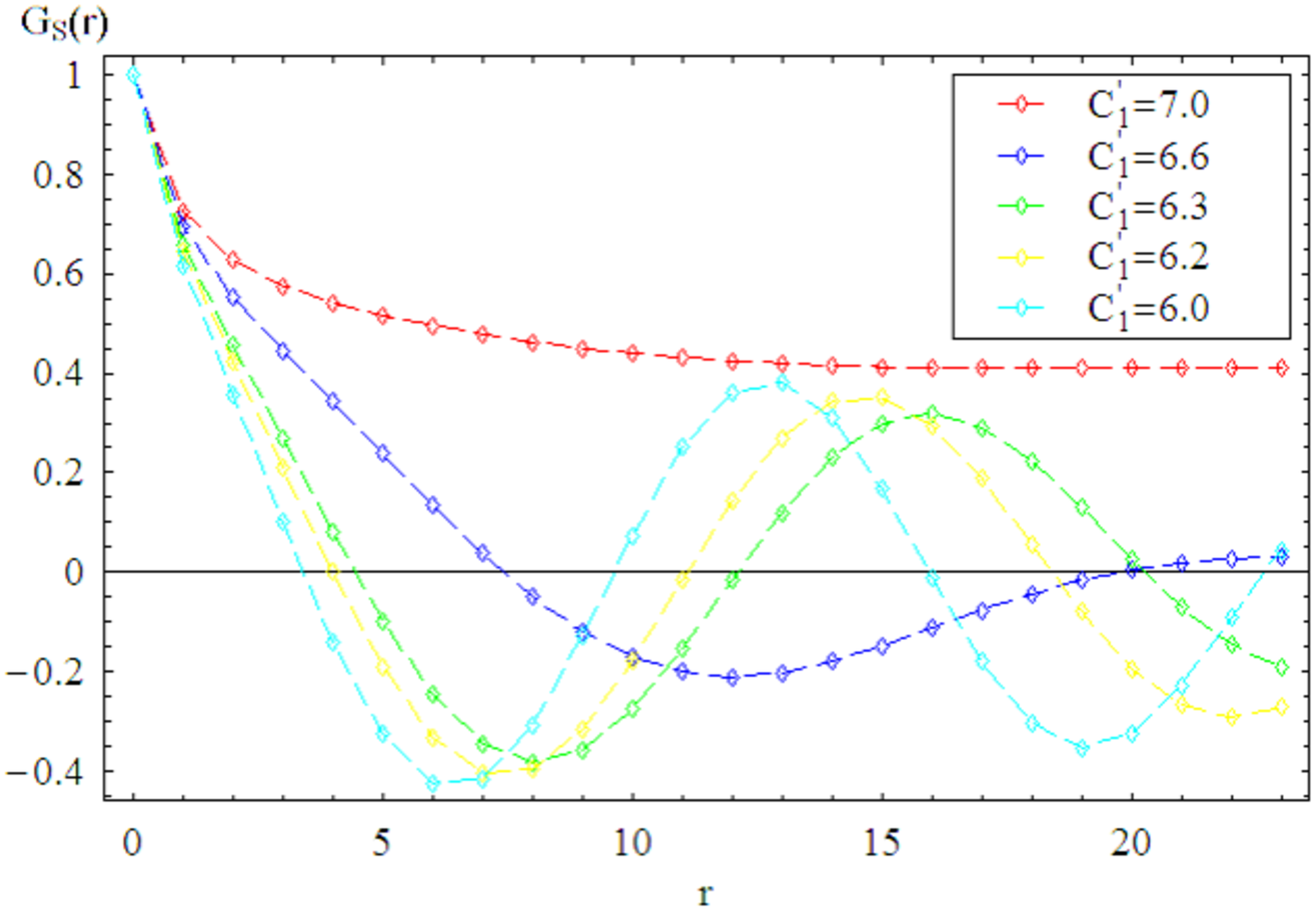}
\includegraphics[scale=.3]{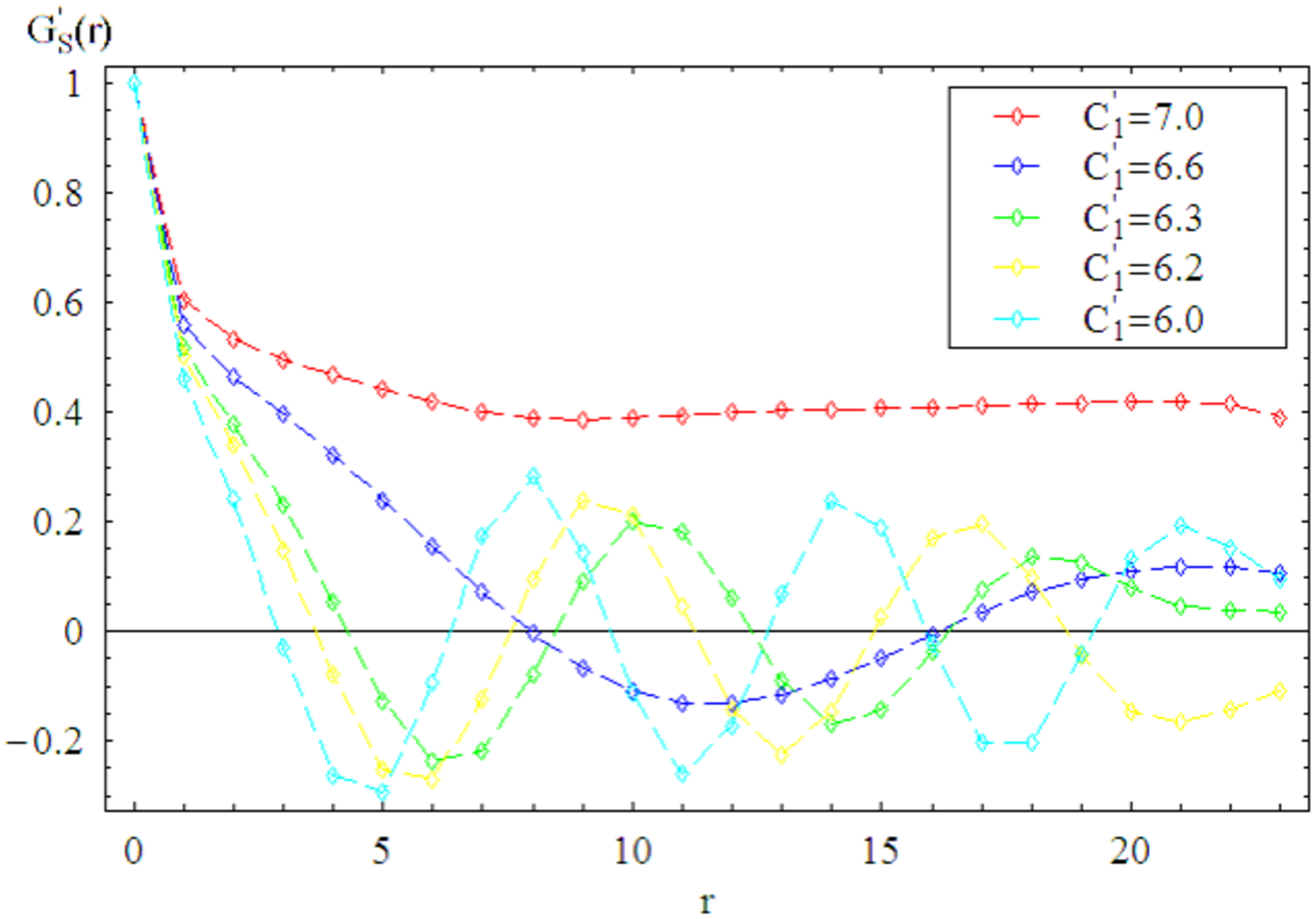}
\end{center}
\caption{Spin correlation in spiral and AF states.}
\label{spin_SP_AF}
\end{figure}

It is also interesting to see how the spin correlation changes
from the spiral to AF phases.
Results in Figs.\ref{spin_SP_AF} show that the spin correlation
gradually changes from the spiral order to AF order.

\begin{figure}[htbp]
\begin{center}
\includegraphics[scale=.4]{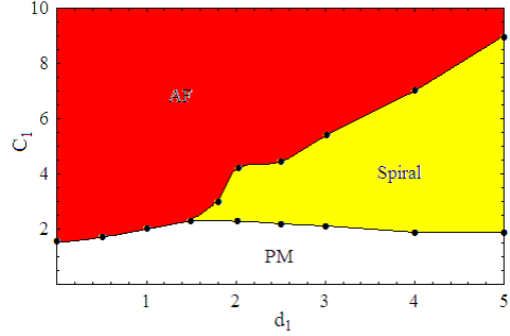}
\end{center}
\caption{Phase diagram of model $S_0$.
There are three phases, AF, PM and spiral phases
as in the model $S_2$.
All phase transition lines are of second order.}
\label{phase2}
\end{figure}
\begin{figure}[htbp]
\begin{center}
\includegraphics[scale=.3]{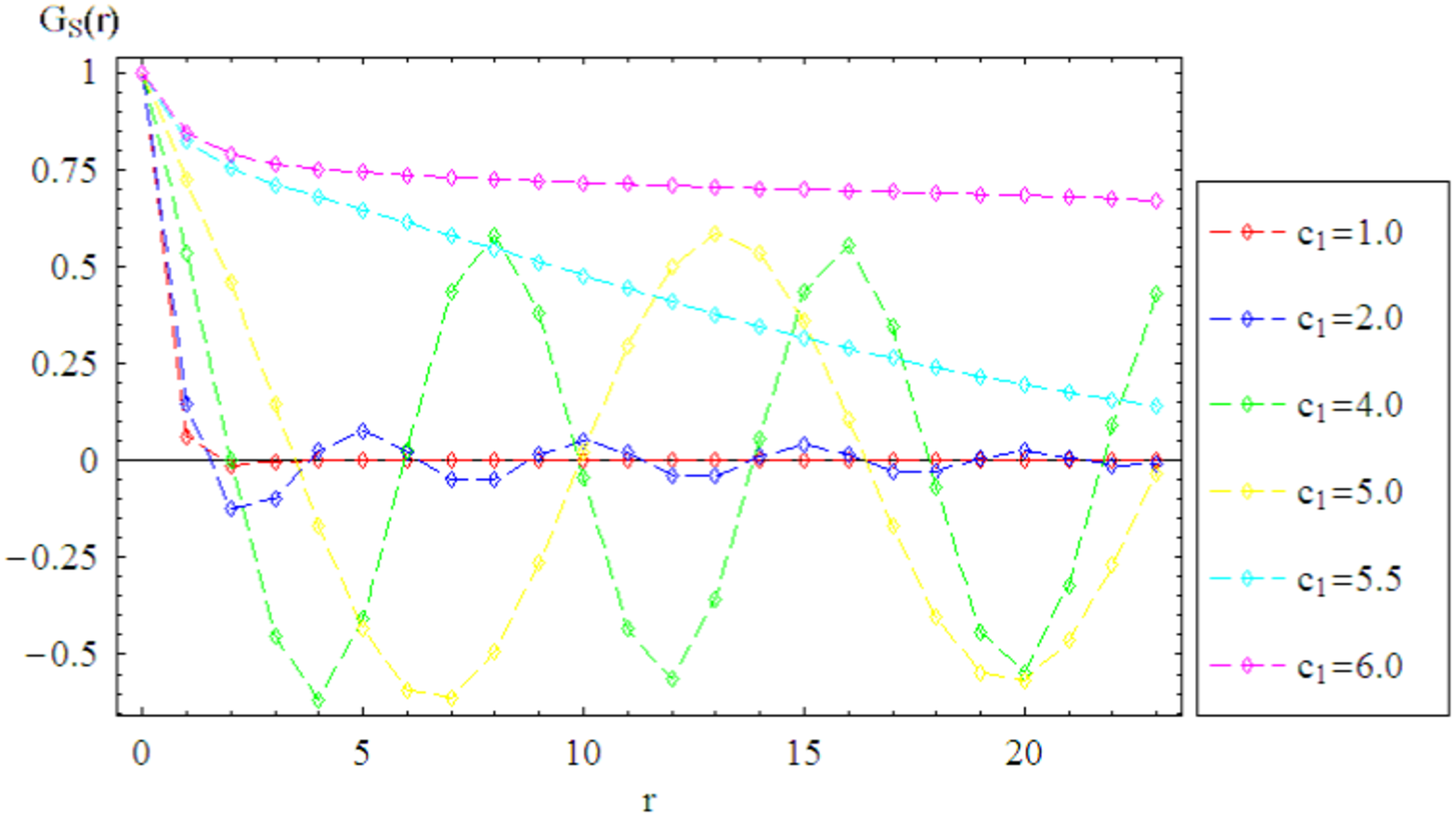}
\includegraphics[scale=.3]{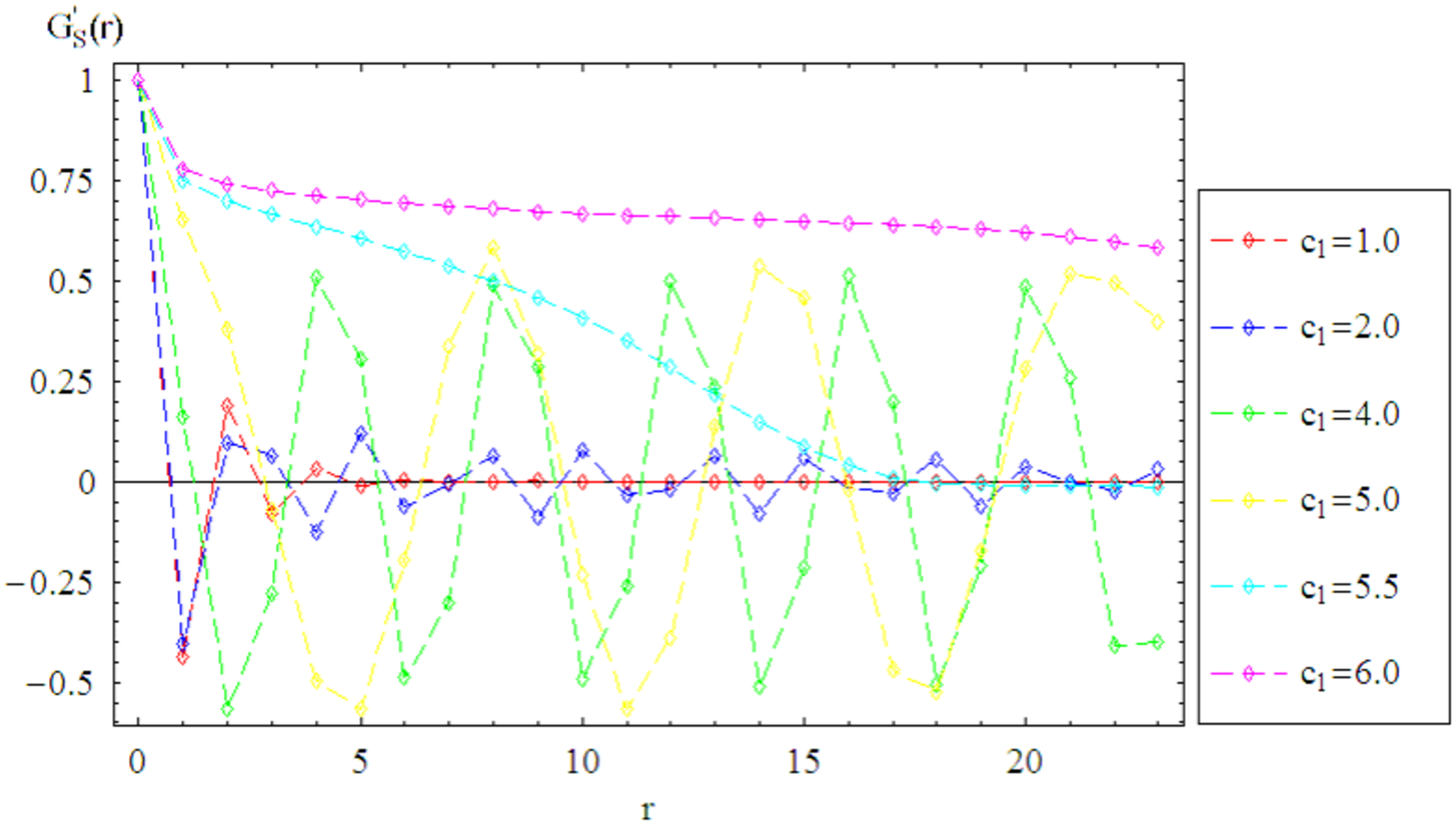}
\end{center}
\caption{Spin correlation functions in model $S_0$ for $d_1=3.0$.}
\label{spincor_S0}
\end{figure}

We also numerically studied the original model $S_0$ in Eq.(\ref{Z2}),
and obtained similar results to those of model $S_2$.
Obtained phase diagram is shown in Fig.\ref{phase2},
and spin correlation functions in Fig.\ref{spincor_S0}.
Phase transitions are of second-order and spin correlation functions
have similar behavior to those of $S_2$. 

Result for the spiral state obtained in the present subsection 
obviously means that the $120^0$-N\'eel state is realized in
each layer for the case of the isotropic triangular case $J=J'$.
This result is in good agreement with the previous study on the 
AF Heisenberg model on isotropic triangular lattice at $T=0$\cite{isotropic}.
On the other hand,
some of the previous study on the anisotropic AF Heisenberg model
on 2D triangular lattice at $T=0$ suggested the existence
of spin-liquid phases\cite{spinliquid}.
However the results obtained in this subsection show that it does
not exist in the present model.
In order to address the possibility of the spin-liquid phase,
we shall study effect of the plaquette term of the emergent
gauge field $U_{x\mu}$ in the following subsection.

\subsection{$c_2>0$ case}

In this subsection, we shall study the effects of the plaquette $c_2$-term
in the action $S_2$.
For the $c_2=0$ case, we found that there exist three phases, 
i.e., the AF, spiral and PM phases.
It is expected that the gauge dynamics in both the AF and spiral phases
is in the Higgs phase as the condensation of the spinon field
$z_x$ suppresses fluctuations of the gauge field $U_{x\mu}$.
Low-energy excitations are gapless spin wave in the both phases.
On the other hand in the PM phase, it is known that the confinement phase
is realized and low-energy excitations are {\em bound state of the spinons}
like a spin-triplet $(\bar{z}_x\vec{\sigma}z_x)$ 
because of the strong fluctuations of $U_{x\mu}$.

As explained in Sec.2, another possible phase in the quantum spin systems
is the spin-liquid phase.
In that phase, there exist no LRO's, whereas low-energy excitation is 
the {\em deconfined spinon} $z_x$.
This means that in the spin liquid only small fluctuations of 
the gauge field $U_{x\mu}$ are
realized and the gauge dynamics is in the Coulomb phase.
Knowledge of the gauge field theory suggests that such a spin-liquid phase
may be realized by turning on the plaquette $c_2$-term because this term
suppresses large fluctuations of the gauge field\cite{kogut}.
In the previous study on the $Z_2$ gauge model of the spiral and
spin-liquid phases\cite{Z2}, we found that the deconfined spin-liquid phase
is realized in the vicinity of the spiral and tilted-dimer states.
In the present paper, we show the results of 
study on the U(1) gauge model $S_2$
in the $c'_1-c_2$ parameter plane with the value of $d_1$ fixed.

\begin{figure}[htbp]
\begin{center}
\includegraphics[scale=.5]{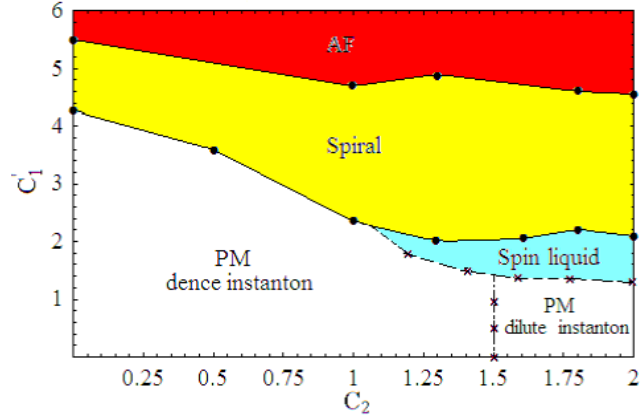}
\end{center}
\caption{Phase diagram in $c_2$-$c'_1$ plane for  $d_1=2.0$.
Spin-liquid phase appears in the vicinity of PM (tilted-dimer)
and spiral phases.}
\label{c2_phase}
\end{figure}

We first show phase diagram obtained by the MC simulations for $d_1=2.0$
in Fig.\ref{c2_phase}.
Reason for choosing this value of $d_1$ is that the spiral and
tilted-dimer states appear as the value of $c'_1$ is varied for
$c_2=0$.  
As shown in Fig.\ref{c2_phase}, there exists a crossover line emanating from
the point $(c_2\simeq 1.5, c_1=0)$ in the vertical direction.
This crossover line separates dense and dilute instanton regions,
whereas the both regions belong to the confinement phase of the U(1)
gauge model in 3D. 
Besides the crossover line, there exist two sharp second-order phase
transition lines emanating from $(c_2=0, c'_1=4.2)$ and $(c_2=0, c'_1=5.5)$,
respectively.
As shown in Fig.\ref{c2_phase}, these are the spiral and AF phase transition lines,
respectively.
We also found another ``transition line" emanating from 
$(c_2\simeq 1.05, c'_1\simeq 2.2)$, 
which we identify as a crossover to the spin-liquid phase.

\begin{figure}[bhtp]
\begin{center}
\includegraphics[scale=.4]{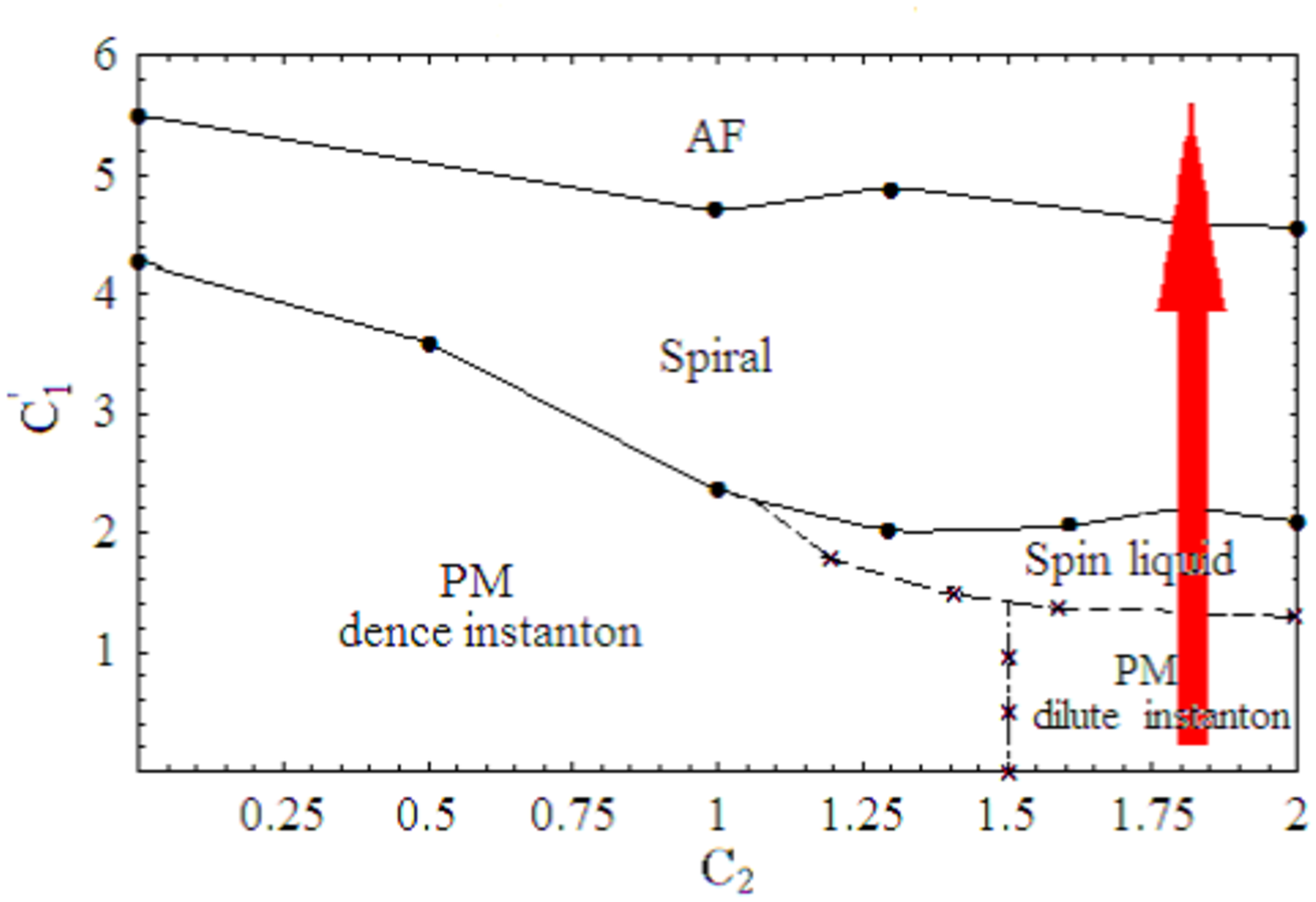}
\includegraphics[scale=.4]{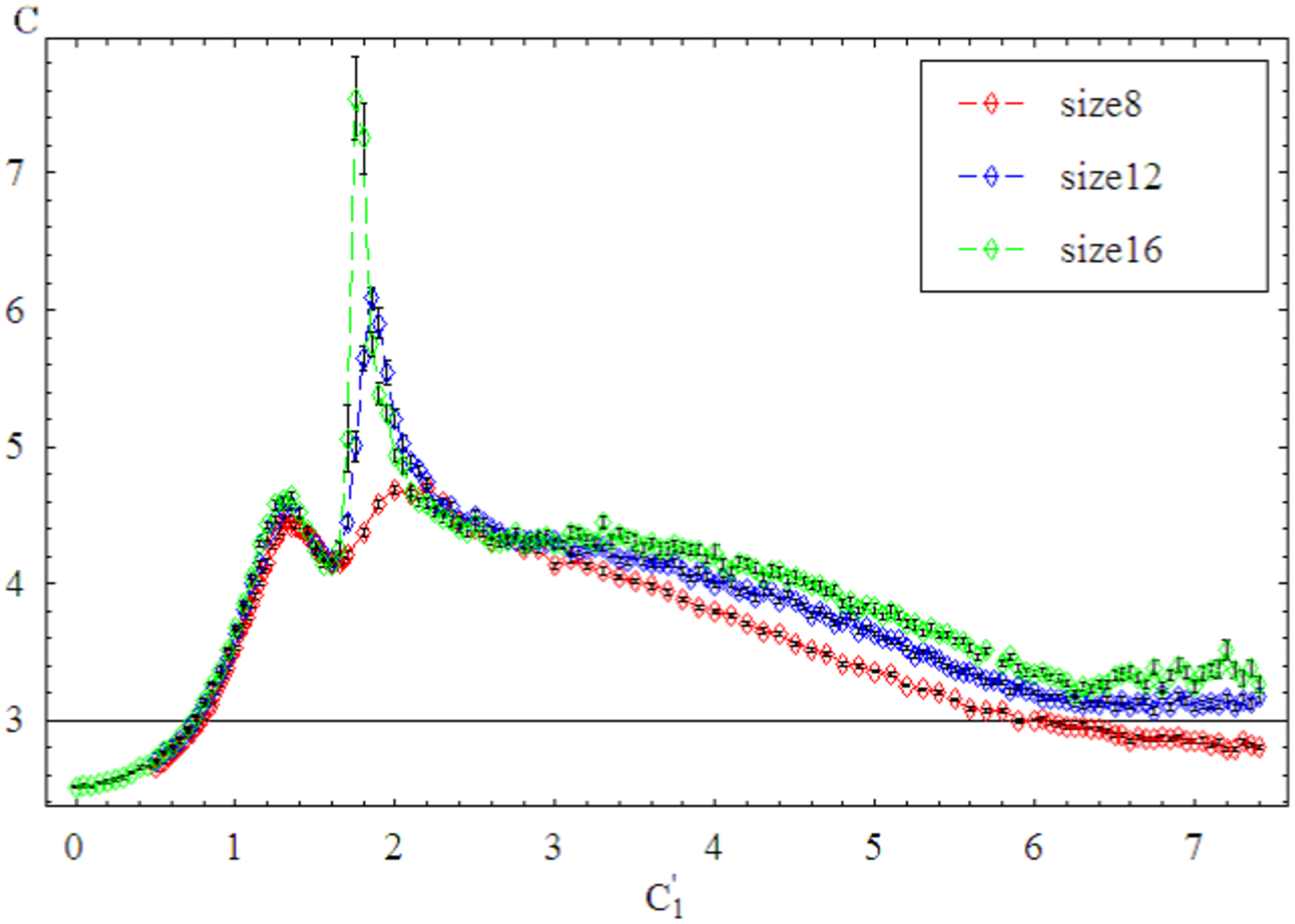}
\end{center}
\caption{$C$ as a function of $c'_1$ for $c_2=1.8$ and $d_1=2.0$.}
\label{C_all_s_liquid}
\end{figure}

\begin{figure}[htbp]
\begin{center}
\includegraphics[scale=.3]{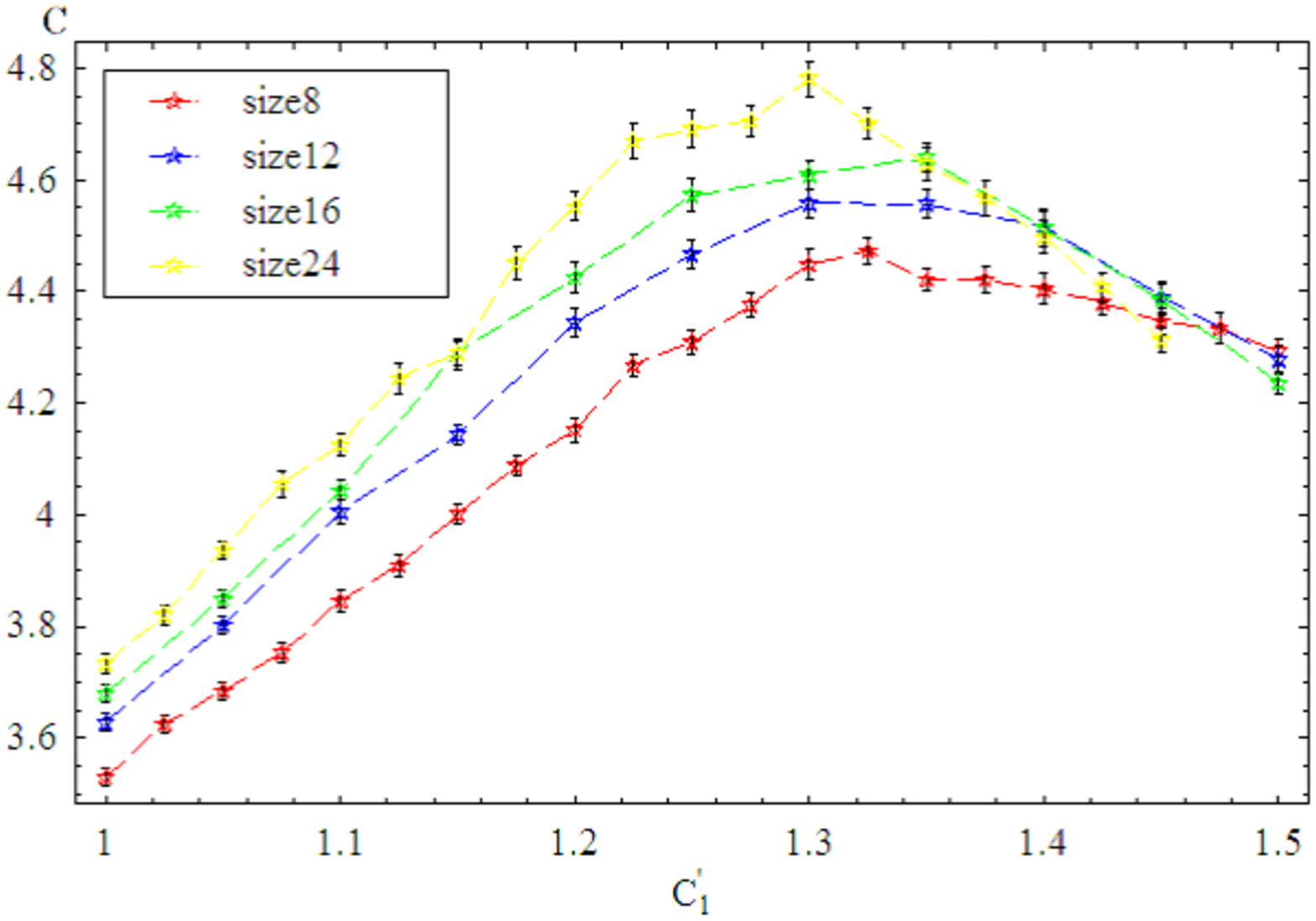}
\end{center}
\caption{Detailed calculation of $C$ as a function of $c'_1$ 
for $c_2=1.8$ and $d_1=2.0$.}
\label{C_all_s_liquid2}
\end{figure}

We show the total specific heat $C$ as a function of 
$c'_1$ for $c_2=1.8$ and $d_1=2.0$
in Fig.\ref{C_all_s_liquid}.
There are two peaks at $c'_1\simeq 1.4$ and $1.8$, and
the second peak at $c'_1\simeq 1.8$ develops as the system size
is increased indicating a second-order phase transition.
Calculation of the spin correlation given later on shows 
that it is the phase transition to the spiral state.
On the other hand, the first peak at $c'_1\simeq 1.4$ does
not develop as the system size is increased.
More detailed calculation is shown in Fig.\ref{C_all_s_liquid2}.
There exists small system-size dependence, but we think that
this size dependence comes from the free-boundary condition that we took
for the calculation.

\begin{figure}[htbp]
\begin{center}
\includegraphics[scale=.25]{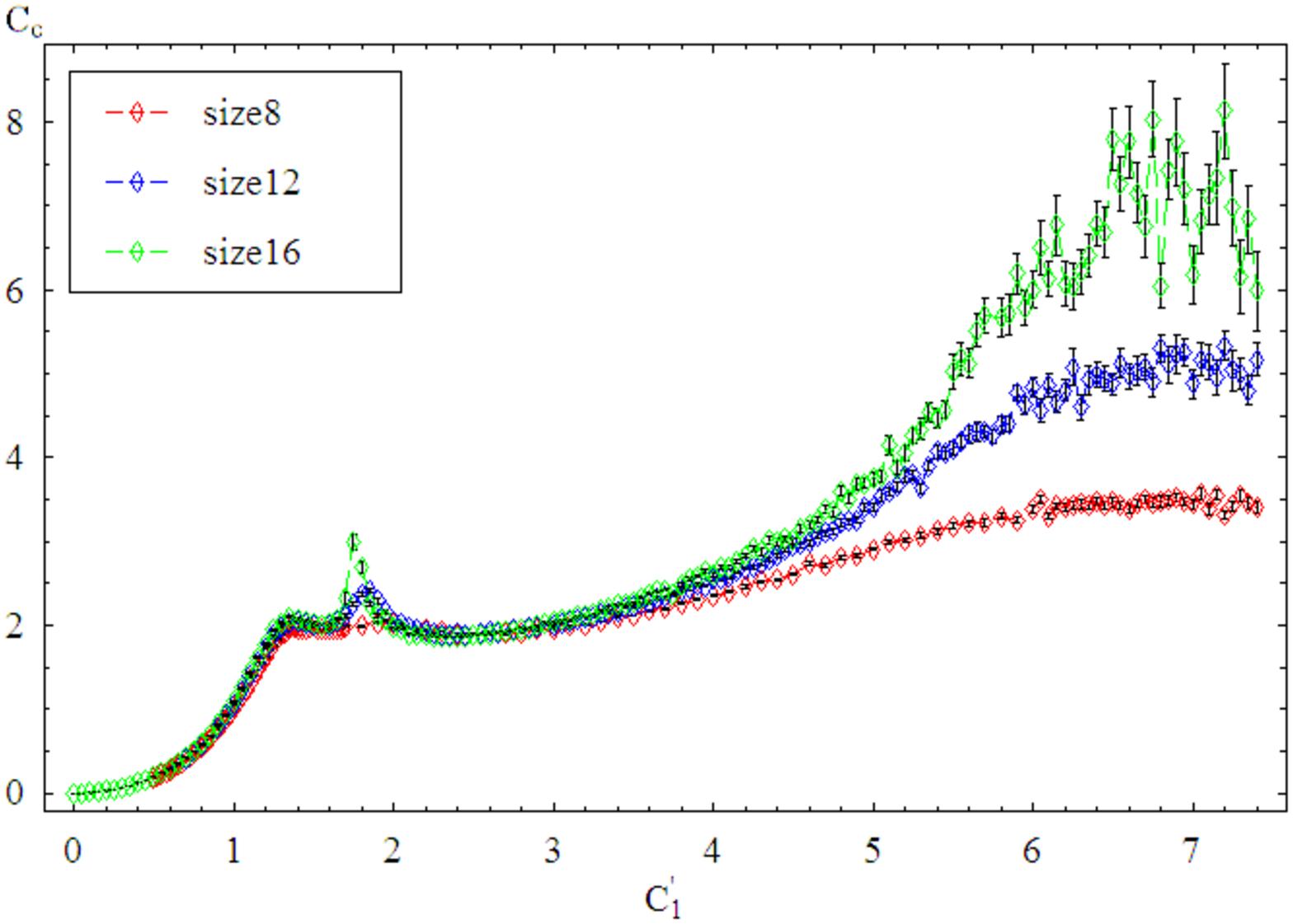}
\includegraphics[scale=.25]{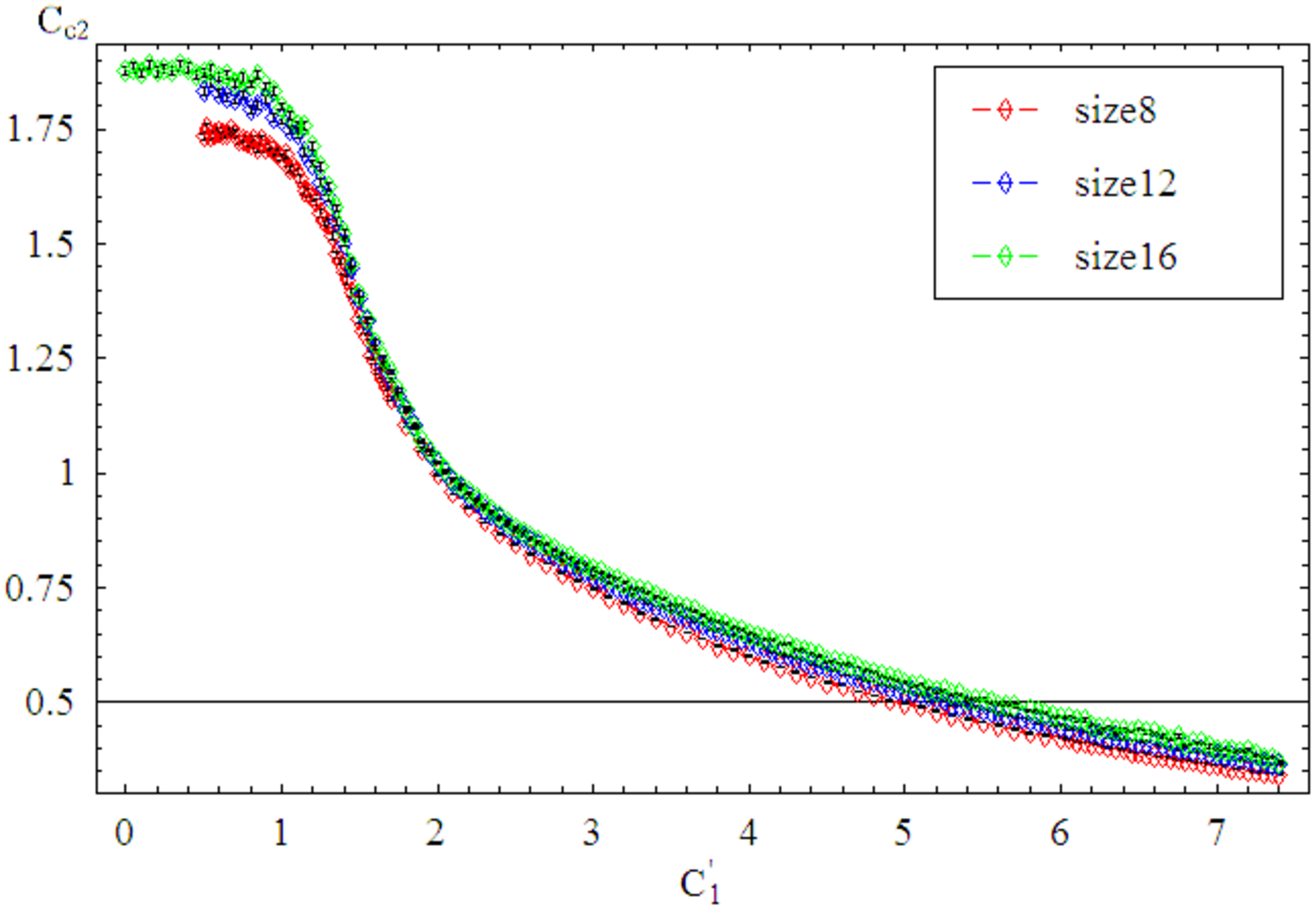}
\includegraphics[scale=.25]{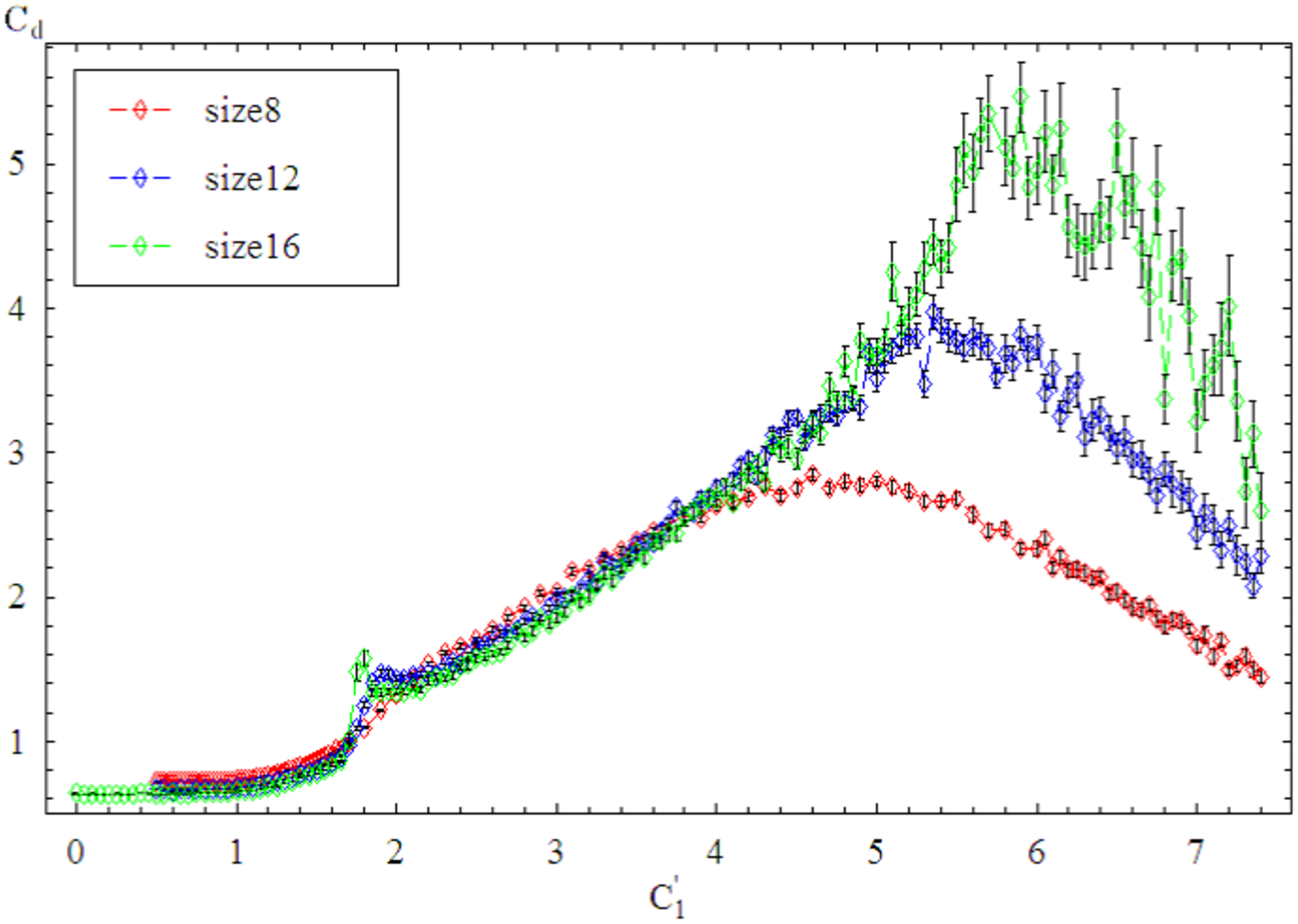}
\end{center}
\caption{Specific heat of each term in action
for $c_2=1.8$ and $d_1=2.0$.}
\label{C_each_s_liquid}
\end{figure}

It is useful to see how specific heat of each term behaves.
See Fig.\ref{C_each_s_liquid}.
The specific heat of the $c_2$-term $C_{c2}$, which is defined similarly to
$C_c$ and $C_d$, is a decreasing function of $c'_1$ and 
changes its behavior at $c'_1\simeq 1.4$.
On the other hand, the specific heats of the $c'_1$-term and $d_1$-term both
have peaks at $c'_1\simeq 1.8$ and $c'_1\simeq 5.5$ and 
these peaks develop as the system size is increased.
This result suggests that there is a second-order phase transition at
$c'_1\simeq 5.5$ besides at $c'_1\simeq 1.8$.

\begin{figure}[htbp]
\begin{center}
\includegraphics[scale=.3]{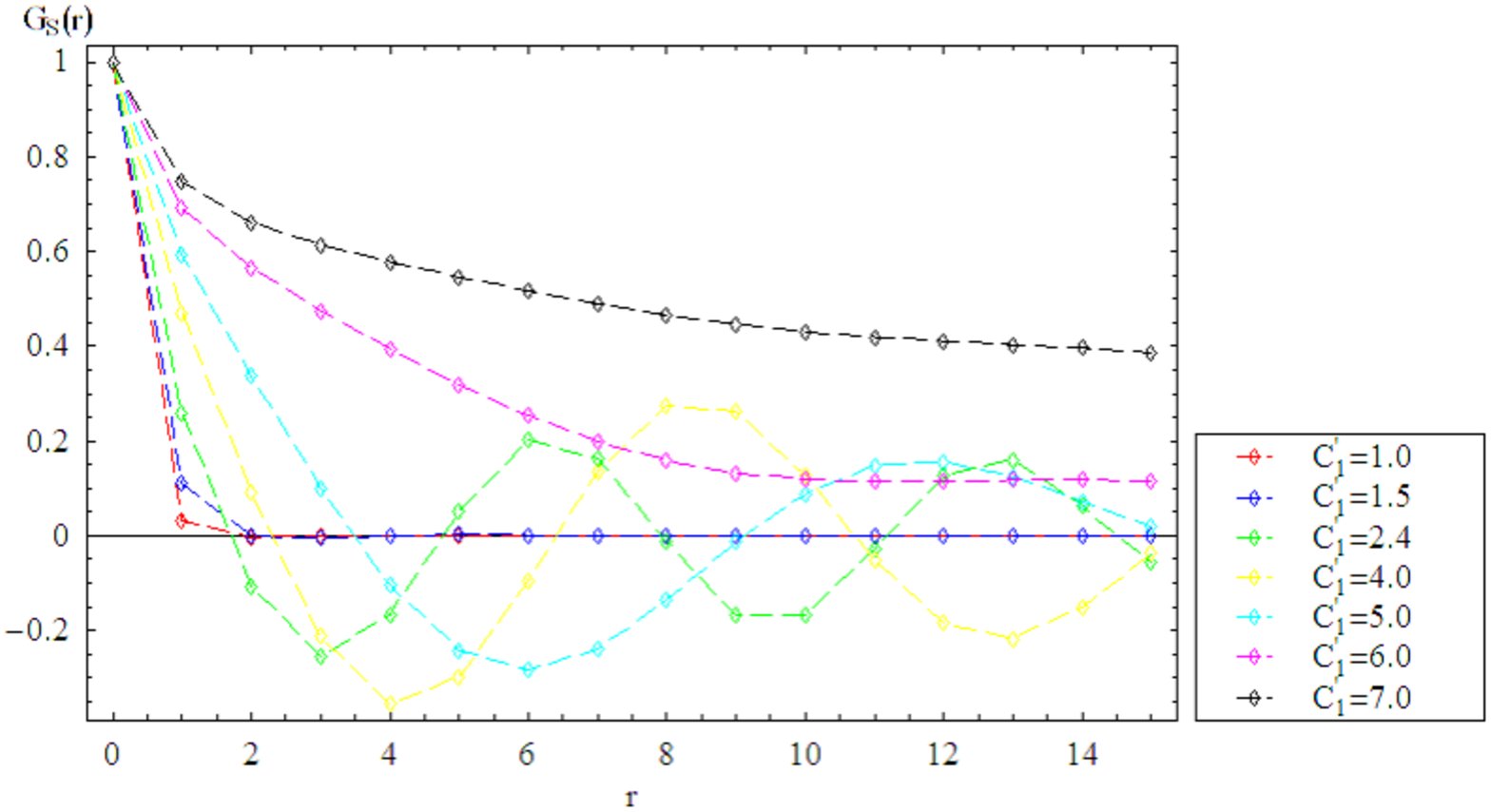}
\end{center}
\caption{Spin correlation function for various values of $c'_1$. 
}
\label{spin_corr_s_liquid}
\end{figure}

It is important to see how the spin correlation function
behaves and verify properties of each phase observed by the measurement of $C$.
Obtained results of the spin correlation for $c_2=1.8,\; d_1=2.0$ are shown in 
Fig.\ref{spin_corr_s_liquid}.
At $c'_1=1.0$ and $1.5$, there exists no LRO, whereas at $c'_1=2.4,\; 4.0$
and $5.0$ the LR spiral order appears.
Furthermore at $c'_1=6.0$, the spin correlation shows the AF LRO.
All the above results verify the phase diagram shown in Fig.\ref{c2_phase}

\begin{figure}[htbp]
\begin{center}
\includegraphics[scale=.3]{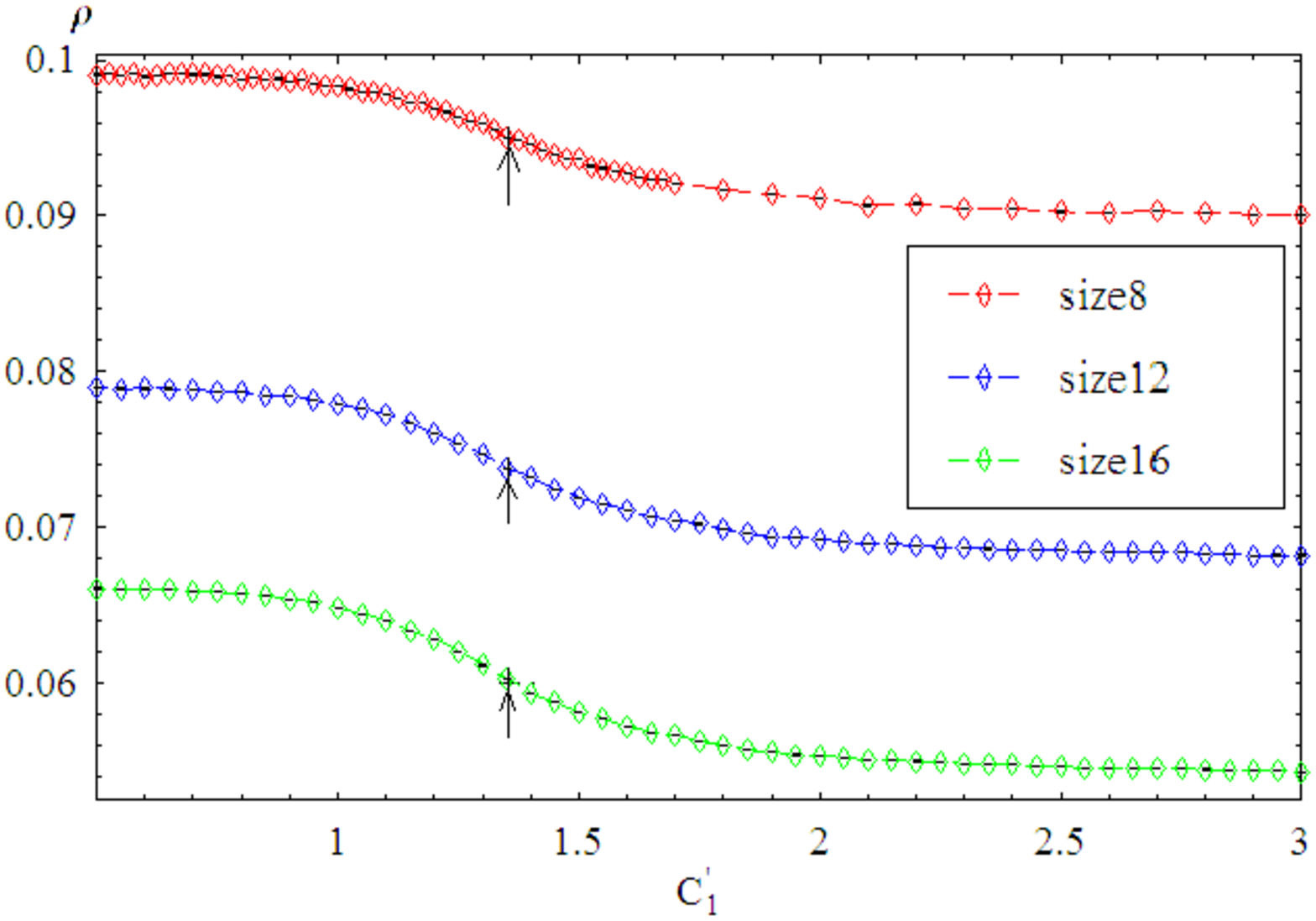}
\end{center}
\caption{Instanton density as a function of $c'_1$. 
Arrows indicate location of crossover observed by calculation
of $C$.
}
\label{instanton_s_liquid}
\end{figure}

In order to investigate the gauge dynamics, it is useful to
study instanton (monopole) density $\rho$, which measures magnitude of
topologically nontrivial fluctuations of the gauge field $U_{x\mu}$.
$\rho(x)$ is defined as follows for the gauge field configuration 
$U_{x,\mu}=e^{i\theta_{x,\mu}}$\cite{Inst-1,CPN-2}.
First we consider the magnetic flux $\Theta_{x,\mu\nu}$ penetrating plaquette
$(x,x+\mu,x+\mu+\nu,x+\nu)$
\begin{eqnarray}
\Theta_{x,\mu\nu}&=&\theta_{x,\mu}+\theta_{x+\mu,\nu}-\theta_{x+\nu,\mu}
-\theta_{x,\nu}, \nonumber \\
 && (-4\pi\le \Theta_{x,\mu\nu}\le 4\pi).
\label{Flux}
\end{eqnarray}
We decompose $\Theta_{x,\mu\nu}$ into its integer part $n_{x,\mu\nu}$,
which represents the Dirac string (vortex line), and the remaining part 
$\tilde{\Theta}_{x,\mu\nu}$,
\begin{equation}
\Theta_{x,\mu\nu}=2\pi n_{x,\mu\nu}+\tilde{\Theta}_{x,\mu\nu}, \;\;
(-\pi\le \tilde{\Theta}_{x,\mu\nu}\le \pi).
\label{Flux2}
\end{equation}
Then instanton density $\rho(x)$ at the cube around the site
$x+{\hat{1} \over 2}+{\hat{2} \over 2}+{\hat{3} \over 2}$ of the
dual lattice is defined as 
\begin{equation}
\begin{split}
\rho(x) &=-{1 \over 2}\sum_{\mu\nu\lambda}\epsilon_{\mu\nu\lambda}
(n_{x+\mu,\nu\lambda}-n_{x,\nu\lambda})  \\
&={1 \over 4\pi}\sum_{\mu\nu\lambda}\epsilon_{\mu\nu\lambda}
(\tilde{\Theta}_{x+\mu,\nu\lambda}-\tilde{\Theta}_{x,\nu\lambda}),
\end{split}
\label{rho}
\end{equation}
where $\epsilon_{\mu\nu\lambda}$ is the antisymmetric tensor.

In Fig.\ref{instanton_s_liquid}, we show the calculation of 
$$
\rho={1 \over L^3}\sum_x |\rho(x)|.
$$
As the gauge dynamics is already in the dilute-instanton region
of the confinement phase for $c'_1=0$, the value of $\rho$ is small,
but it decreases at $c'_1\simeq 1.4$, and its behavior becomes
clear as the system size is increased.
This result indicates that the region between two peaks at $c'_1=1.4$
and $1.8$ corresponds to the ``deconfined Coulomb phase".
In this phase, the gapless gauge boson $\theta_{x\mu}$ appears
as a low-energy excitation besides the deconfined spinons.

The global phase diagram in Fig.\ref{c2_phase} is consistent
with that of the $Z_2$ spin-liquid model obtained in Ref.\cite{Z2}.
In the $Z_2$ model, however, the phase transition from spiral phase to 
PM phase is of first order and there exists sharp phase boundary 
between spin-liquid and PM phases.
Anyway, results obtained in this paper support discussions
in term of $Z_2$ models for frustrated quantum AF magnets at $T=0$.

\section{Summary}
\setcounter{equation}{0}

In this paper we have studied phase structure of the AF spin model
on the layered anisotropic triangular lattice.
We used the Schwinger bosons for represent ting quantum spins
and also employed the coherent-state path integral methods.
We focused on finite-$T$ phase diagram and investigated it
by means of the MC simulations.
We calculated the internal energy, specific heat and spin correlation
functions.
In the absence of the $c_2$-term, we found that 
there exist three phases, i.e., AF, PM and spiral phases.
All phase transitions between them are of second-order. 

\begin{figure}[htbp]
\begin{center}
\includegraphics[scale=.4]{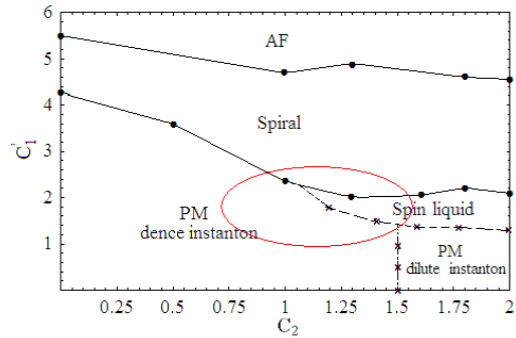}
\end{center}
\caption{Encircled region of the phase diagram was studied in the 
previous paper in terms of gauge theories with local $Z_2$ gauge
symmetry\cite{Z2}. The results obtained in Ref.\cite{Z2} and 
the present paper show that there exist the PM-tilted-dimer,
spiral and deconfined spin-liquid phases in the AF magnets
on triangular lattice at low $T$.
}
\label{phase_z2}
\end{figure}

Then we turned on the $c_2$-term and investigated if the deconfined
spin-liquid phase appears.
Calculations of the specific heat and instanton density shows that
there exists deconfined spin liquid in the vicinity of the spiral and
tilted dimer states.
This result is in good agreement with the results of our previous study\cite{Z2}
in which we assumed a short-range spiral order and focused on 
the region in the phase diagram shown in Fig.\ref{phase_z2}.
However, the present study indicates that there is not sharp phase
boundary between the tilted-dimer state and spin liquid, i.e.,
this ``transition" is a crossover.
At very lower $T$, it is possible that this crossover changes to a genuine
phase transition as the imaginary time plays a role of another dimension.

It is very interesting to study how hole doping changes the observed
phase diagrams and how doped holes behaves in various magnetic phases.
This problem is under study and we hope that we shall report the results
in a future publication.

\bigskip
\acknowledgments
This work was partially supported by Grant-in-Aid
for Scientific Research from Japan Society for the 
Promotion of Science under Grant No.20540264.


\end{document}